\documentclass[twocolumn,pra,preprintnumbers,amsmath,amssymb,superscriptaddress]{revtex4}
\usepackage{graphicx}
\pdfoutput=1

\newcommand{\bea}{\begin{eqnarray}}
\newcommand{\eea}{\end{eqnarray}}
\newcommand{\p}{\partial}

\begin{document}

\preprint{YITP-12-104}

\title{Stripe Instabilities of Geometries with Hyperscaling Violation}

\date{\today}

\author{Norihiro {\sc Iizuka}}\email[]{iizuka@yukawa.kyoto-u.ac.jp}
\affiliation{%
{\it Yukawa Institute for Theoretical Physics, 
Kyoto University, Kyoto 606-8502, JAPAN}}

\author{Kengo {\sc Maeda}}\email[]{maeda302@sic.shibaura-it.ac.jp}
\affiliation{%
{\it Faculty of Engineering,
Shibaura Institute of Technology, Saitama 330-8570, JAPAN}}

\begin{abstract}
We study the dynamical stripe instabilities on the geometries with hyperscaling violation in the IR, which asymptotically approach AdS$_4$ in the UV. The instabilities break the translational invariance spontaneously and are induced by the axion term $\sim a F \wedge F$ in the bulk action. We first study the perturbation equations in the probe limit, and find that there is a strong correlation between the stripe instabilities caused by the axion term and parameters of the theories which determine the IR hyperscaling violation. Contrary to the IR AdS$_2$ case, the effect of the axion term for the stripe instabilities can be enhanced/suppressed at low temperature depending on the parameters. For a certain one-parameter family of the hyperscaling violation, we find the onset of the stripe instability analytically in the axion coupling tuned model. For more generic parameter range of hyperscaling violation, we study the instability onset by searching for the zero mode numerically on the full geometries. We also argue that quite analogous results hold, after taking into account the graviton fluctuation, {\it i.e.,} beyond the probe limit. 
\end{abstract}

\maketitle

\tableofcontents


\noindent

\section{Introduction}\label{sec:intro}

Holography is an extremely useful tool to understand the quantum field theory in the 
strongly coupled limit \cite{Maldacena:1997re,Witten:1998qj,Gubser:1998bc}. 
This motivates us to apply the holographic technique to   
interesting subjects 
such as 
QCD 
and condensed matter physics,  
and see what we can learn about   
real world physics from holography. 

In studying the quantum field theories for QCD or condensed matter physics, 
one of the most important things is understanding the infrared (or vacuum) structure. 
For example, in many of the recent studies of the non-Fermi liquid in holographic setting,  
the most interesting nature is originated from its IR behavior, {\it i.e.,} the 
bulk black brane near horizon geometries 
\cite{Liu:2009dm, Faulkner:2009wj, Cubrovic:2009ye}.   
This is natural from the trivial fact that Fermi surface is the low energy nature of the systems. 
In the holographic superconductor case \cite{Gubser:2008px, Hartnoll:2008vx, Hartnoll:2008kx, Herzog:2009xv, Horowitz:2010gk, Horowitz:2009ij}, with or without various lattice effects \cite{Horowitz:2012ky, Horowitz:2012gs, Iizuka:2012dk}, 
the most interesting features are their symmetry breaking pattern in the IR, {\it i.e.,} 
in the near horizons of hairy 
black branes.  
Another interesting aspects of black brane geometries for the condensed matter application is the 
existence of flux at the horizon, since this gives the violation of the Luttinger theorem 
for the field theory duals 
\cite{Hartnoll:2010xj, Sachdev:2011ze, Huijse:2011ef, Huijse:2011hp, Iqbal:2011bf, Hashimoto:2012ti} 
and this resembles the  
fractionalized Fermi-liquids \cite{Senthil:2003,Sachdev:2010um,Huijse:2011hp}. 
Therefore it gives the possibly interesting dual of them. 
Quantum criticality is another fine example where IR dominates the physics. 
In these ways, we are especially interested in the IR behavior  
of the black brane near horizon geometries in holographic condensed matter, 
and it is quite interesting to understand their various dynamics from the geometry side. 
See also 
\cite{Iizuka:2012iv, Iizuka:2012pn} for recent development of the classification of the black brane geometries which admits homogeneous but anisotropic geometries.  

One of the most interesting geometries in this view point which people actively studied recently 
is Lifshitz geometry \cite{lifsol1,lifsol2,lifsol3} and so-called geometry with hyperscaling violation \cite{hv1,Iizuka:2011hg,hv31,hv3,hv32,hv4,hv5,hv6,Edalati:2012tc,Bueno:2012vx}. 
These geometries are interesting since they are, if realized in the IR, dual to the field theories 
which break the Lorentz-invariance but respect spatial rotation and translational invariance in the IR.  
IR Lifshitz and hyperscaling violating geometries can be realized as the near horizon geometries  
of some black brane solutions, and they can be obtained, for example, in theories where  
dilaton has run-away behavior governed by the exponential potentials \cite{lifsol3, hv1, Iizuka:2011hg}.
 
Another interesting nature of these geometries is that these geometries 
 do not admit large entropy unlike 
Reissner-Nordstrom black branes \footnote{One can also construct black brane solutions which 
do not have large entropy by admitting spatial anisotropy, see for example, \cite{Iizuka:2012wt}}. 
On the boundary field theory side, this means that it has vanishing entropy density at the 
zero temperature limit, which is more natural from thermodynamical view point.   
In addition,  
rich behavior of the hyperscaling violating geometries allows more exotic behavior  
for the fermion correlators. For examples, it gives various $\omega$-dependence for the non-Fermi liquid decay-rates, as shown in \cite{Iizuka:2011hg}. 

In the meantime, recently a lot of recent progress was made in 
understanding the translational symmetry breaking 
in the holographic setting \cite{ooguri,gauntlett, Donosstripes, 
holstripes1, holstripes2, holstripes3, Rozali, DGnew}. 
The symmetry breaking 
can be induced by the  
axion term $a F \wedge F$ 
in 4d gravity (where $a$ is neutral pseudo-scalar field, and we call it axion in this paper), and 
can occur both spontaneously and by the source term. 
In many of the situations, the analysis is mainly done for the Reissner-Nordstrom black brane 
which admits AdS$_2$ geometries in the IR. 
Given this, it is very natural to study if above IR Lifshitz or geometries with hyperscaling violation 
can survive after we take into account such axion effect and see if it induces the 
translational symmetry breaking. 
Furthermore, since this instability changes the IR nature of the geometries, 
studying this instability on the 
hyperscaling violating geometries is by itself interesting questions as general relativity problem. 
In this paper, 
we study the instability on the IR hyperscaling violating geometries, focusing on the 
onset of the stripe instability. 

The organization of this paper is as follows; 
We first review the geometries with hyperscaling violation in the 4d Einstein-Maxwell-dilaton-axion 
system in \S II. 
The geometries with hyperscaling violation emerge in the IR, {\it i.e.,} in the near horizon limit 
of the full solutions which  
approach AdS$_4$ in the UV asymptotically. 
In \S III, we study the perturbation 
on these 
geometries in the probe limit. 
We first analyze the perturbation 
focusing on the IR geometries 
at small nonzero temperature in \S \ref{section3B}, and at zero temperature limit \S \ref{section3C1}. 
These studies allow us to find the dependence of the 
stripe instabilities 
on the parameters of the theories which determine the IR hyperscaling violation. 
Given this, we identify the instability onset 
on a certain one-parameter family of the hyperscaling violating geometries, 
in \S \ref{section3C2}. 
In \S \ref{section3D}, we search for the 
zero mode on the full geometries for more generic parameters numerically. 
In \S IV, we study the perturbation equations without taking the probe limit, 
and see that essentially the same results hold compared to the 
probe limit case. We end with a summary and discussion in \S V.

\section{Einstein-Maxwell-dilaton-axion system} 
\subsection{The set-up} 
\label{section2A1}

The action we consider is Einstein-Maxwell theory coupled to a dilaton-axion, given by
\bea
\label{actionofourmodel}
S=\int d^4x\sqrt{-g}\bigl(R  -2(\nabla\phi)^2 - 2  e^{2 \xi \phi} (\nabla a)^2 - V(\phi) \nonumber \\ - f(\phi) F_{\mu\nu} F^{\mu\nu} -\theta(a)F_{\mu\nu} \tilde{F}^{\mu\nu} \Bigr) \,. \quad
\eea 
Here, 
$\tilde F^{\mu\nu} \equiv \frac12 \epsilon^{\mu\nu\rho\kappa} F_{\rho \kappa}$, 
and $\epsilon^{\mu\nu\rho\kappa}$ has a factor of $1/\sqrt{-g}$ in its definition such that 
axion term $\theta(a)F_{\mu\nu} \tilde{F}^{\mu\nu}$
is independent of the metric. We take the convention that $\epsilon_{trxy} > 0$. 

As an explicit example, in this paper we consider 
\bea
\label{explicitexamples}
& f(\phi) = e^{2 \alpha \phi} 
\,, \quad 
V(\phi) =  2 V_0 \cosh{2 \delta \phi}  \,, \quad  
 \theta (a) = c_1 a \,, \quad  \,.
\eea
for the explicit stability/symmetry breaking by taking various real parameters $\alpha$, $\delta$, $c_1$, and $V_0$, 
but we take $V_0 < 0$.  
Note that $V(\phi) \to  V_0 e^{2 \delta \phi} $ in the $\delta \, \phi \to \infty$ limit.

The Einstein equation  in trace reversed form becomes 
\bea
\label{tracereversedeinstein}
&& R_{\mu\nu} - 2 \partial_\mu \phi \partial_\nu \phi - 2 e^{2 \xi \phi}  \partial_\mu a \partial_\nu a  \nonumber \\ && 
= f(\phi)\left( 2 F_{\mu \lambda} F_\nu^{\,\,\lambda} - \frac12 g_{\mu\nu} F^2 \right) + \frac12 g_{\mu\nu} V(\phi, a)\,,
\eea
which is irrelevant to the axionic $\theta(a) F \wedge F$ term. 
The equations of motion for dilaton and axion are 
\bea
\label{dilatonEOM} && 
\frac{4}{\sqrt{-g}} \partial_\mu (\sqrt{-g} g^{\mu\nu} \partial_\nu \phi) \nonumber \\
&&  \quad \quad = \partial_\phi  V(\phi, a) + (\partial_\phi f(\phi) ) F^2 + 4 \xi e^{2 \xi \phi} (\nabla a)^2   \,,  \quad \\
&&  \label{axionEOM}
 \frac{4}{\sqrt{-g}}  \partial_\mu (\sqrt{-g} g^{\mu\nu} \partial_\nu a )  
 =   ( \partial_a\theta(a) )  F \tilde F  \,, 
\eea
and for the gauge field, we have 
\bea
\label{EOMforgaugeboson}
\partial_\mu \left( \sqrt{-g} \left( f(\phi) F^{\mu\nu} + \theta(a) \tilde F^{\mu\nu} \right) \right) = 0 \,,
\eea
with the Bianchi identity 
\bea
\label{Bianchi}
\partial_\mu F_{\nu\lambda} + \partial_\nu F_{\lambda\mu} + \partial_\lambda F_{\mu\nu} = 0 \,.
\eea

\subsection{Background solutions} \subsubsection{IR Hyperscaling violating geometries}
\label{section2B1}

First we review the background  
hyperscaling violating geometries \cite{hv1, Iizuka:2011hg}, 
on which we later add the small fluctuation to study its stability. 
We consider the background geometries where all quantities are functions of only the radial direction $r$. 
By using the $r$ coordinate re-definition, we make the metric in the form 
\bea
\label{metricansatz}
ds^2 = -\tilde a(r)^2 dt^2 + \frac{dr^2}{\tilde a(r)^2} + b(r)^2 (dx^2 + dy^2)
\eea
and equations of motion and the Bianchi identity are solved by 
\bea
\label{fluxansatz}
F = F_{tr}(r) dt \wedge dr  \,, \quad
F_{tr}(r) = \frac{Q_e}{b^2 f(\phi)} \,.  
\eea
In the IR limit, we choose $\delta \, \phi \to \infty$, such that $V(\phi) \to V_0 e^{2 \delta \phi}$. 
Then with the approximate potential  $V(\phi) = V_0 e^{2 \delta \phi}$ in IR,  
we can obtain the solutions as given by \cite{Iizuka:2011hg},
\bea
\label{finiteTtildear}
 && \tilde a(r)^2 = C_a^2 r^{2 \gamma} \left(1 -  \left( \frac{r_h}{r} \right)^{2 \beta + 2 \gamma - 1}  \right)   \,, \quad \\
 \label{finiteTb}
&& b(r)^2 = r^{2 \beta}  \,, \quad  
\phi(r) = k \log r \,,
\eea
where 
\bea
\label{case1}
&& \beta = { (\alpha+\delta)^2 \over 4 + (\alpha+\delta)^2}   \,, \quad
\gamma = 1 -{ 2 \delta (\alpha+\delta) \over 4 + (\alpha+\delta)^2}   \,, \\
\label{case2}
&& k = - { 2 (\alpha+\delta) \over 4 + (\alpha+\delta)^2}  \,, \,\,  Q_e^2 = -V_0{ 2 - \delta (\alpha+\delta) \over 2 \left( 2 + \alpha (\alpha+\delta) \right)}   , \,\,\quad \\
\label{case3}
&& C_a^2  = -V_0 {\left( 4 + (\alpha+\delta)^2 \right)^2 \over 2 \left(2 + \alpha (\alpha+\delta) \right) \left( 4 + (3 \alpha-\delta) (\alpha+\delta) \right)} , \quad \quad
\eea
and axion field $a$ is set to be zero, $a = 0$ \footnote{This solution is obtained 
under the assumption $Q_e \neq 0$. For the case $Q_e = 0$, the derivation of this solution in 
\cite{Iizuka:2011hg}
breaks down since we do not have to require the condition eq.~(2.19) in \cite{Iizuka:2011hg}. Actually 
AdS$_4$ is such a limit, where we have $\beta = 1$, $\gamma = 1$, $k = 0$, $\delta = 0$, $C_a^2 = -{V_0/6}$ and $Q_e = 0$ with arbitrary choice of $\alpha$.}. 
There are parameter space which satisfies $Q_e^2 > 0$, $C_a^2 > 0$, $\gamma > 0$, $\gamma - \beta > 0$, $\beta > 0$ and $\delta \, k < 0$.  
%
%
The condition $\gamma > 0$ arises from the requirement that $g_{tt}$ vanishes at the horizon in zero temperature limit, and $\gamma -\beta > 0$ arises from the null energy condition at the $r_h = 0$ case. 
$\beta > 0$ is automatically satisfied and $\delta \, k < 0$ is for $\delta \, \phi \to \infty$ at $r \to 0$, which gives $\delta ( \alpha + \delta) > 0$. 

The temperature $T$ of the system is given in terms of $r_h$ as 
\bea
\label{Thinregardstorh}
T =\frac{(2\gamma+2\beta-1) {C_a}^2}{4\pi} \, r_h^{2\gamma-1} \,. 
\eea 
So we require $\gamma > 1/2$ so that $r_h \to 0$ at $T \to 0$. 
Then, with $\beta > 0$, this geometry is thermodynamically stable since 
 the horizon area vanishes at the zero temperature limit and dual theories satisfy the 3rd law of the thermodynamics.

In zero temperature limit, the metric reduces to 
\bea
\tilde a(r) = C_a r^\gamma \,, \quad b(r)  = r^\beta \,, 
\eea 
therefore the horizon locates at $r=0$. 
%
%
This zero temperature metric is sometimes written as 
\bea
&& ds_d^2 =  \tilde r^{- \frac{2}{d} \left( d - \theta \right)} \left( - \tilde r^{-2 (z-1)} dt^2 +l^2 {d\tilde r^2}+ dx^2 + dy^2\right)  \,, \nonumber 
\\
\label{thedefinitionofzandthetaforhyper}
&& z = 1 - \frac{\gamma - \beta}{1 - \gamma - \beta} \,, 
\quad  \theta = 4 \left( \frac{1 - \gamma}{1 - \gamma - \beta}  \right) \,, \\
&& 
l^2 = \frac{1}{(1 - \gamma - \beta)^2 C_a^2} \,, 
\eea
by the coordinate transformation $\tilde r \equiv r^{1 - \gamma - \beta}$, where $d=4$. 

These are the so-called ``geometries with hyperscaling violation''.    
The parameter $z$ is called dynamical critical exponent and 
$\theta$ is called Òhyperscaling violationÓ
parameter. 
For generic values of $\alpha$ and $\delta$, we have $\gamma \neq 1$ and $\theta \neq 0$.  
%
If $\gamma = 1$, and equivalently, $\theta = 0$, this IR metric has additional scale invariance symmetry, 
therefore $\theta$ characterizes the ``deviation'' from the scale invariant limit.  
Similarly, if $z = 1$,  then time and spatial coordinate transform equivalently, therefore 
$z$ characterizes the ``deviation''  from the relativistic limit.

Before we continue, let us pose to check the number of the parameters of our system. 
Our model, given by eq.~(\ref{actionofourmodel}) and (\ref{explicitexamples}), has 5 real parameters, $\alpha$, $\delta$, $V_0$, $\xi$, $c_1$.  
However $V_0$ is a parameter to set the scale (of asymptotic AdS$_4$), so we can set $V_0 = -1$, 
without loss of generality. Furthermore, since our background has zero VEV for the axion field, $a = 0$, 
parameters $\xi$ and $c_1$ are irrelevant for the background. In this way, the background is 
characterized by two real parameters $\alpha$ and $\delta$ only. Especially in the IR, these two parameters 
determine the hyperscaling violation $\theta$ and $z$, or equivalently $\beta$ and $\gamma$. 

We will now review the full solutions which interpolate these IR $\beta$, $\gamma$ geometries 
 with hyperscaling violation to UV AdS$_4$ asymptotically. 
   

\subsubsection{
Full solutions interpolating IR hyperscaling violating geometries   
to UV AdS$_4$} 
\label{section2B2}

The solution we reviewed is the IR limit of the full solution which asymptotically approaches AdS$_4$ in UV. 
In order to obtain the full solution which connects the IR to UV, we perturb the IR solution by $O(\epsilon)$ 
and numerically follow the evolution of it. To obtain 
asymptotic AdS$_4$ with appropriate boundary conditions, 
we tune $\epsilon$, where $\epsilon$ is a small parameter. 
In the zero temperature case, 
we set \cite{Iizuka:2011hg}
\bea
\label{deviation1}
\tilde a(r) &=& C_a r^\gamma \left( 1 + \epsilon d_1 r^\nu \right) \,,\\
\label{deviation2}
b(r) &=& r^\beta \left( 1 + \epsilon d_2 r^\nu \right) \,, \\
\phi(r) &=& k \log r + \epsilon d_3 r^\nu \,,
\eea
and the perturbation equations of motion are all satisfied up to the $O(\epsilon)$ by setting 
\bea
&& \nu_1 = \nu_2  = \nu \,,   \quad 
d_2 =  \frac{ B_1 }{B_2}  d_1  \,, \\  
&& d_3 = \frac{ 4 (-1 + \nu) + (\alpha + \delta)^2 (1 + \nu) }{4 (\alpha + 
   \delta)} d_2  \,,   \\
&&  
\nu = \frac{2 (\delta  (\alpha +\delta
   )+2)}{(\alpha +\delta )^2+4} -\frac{3}{2} + \frac{
   A   }{2  \left((\alpha +\delta )^2+4\right)^2} \,,  \\
&& A =   \Bigl(   \left[  4+ (3 \alpha -\delta ) (\alpha   +\delta )\right] 
    \left[  (\alpha +\delta )^2+4   \right]^2  \nonumber \\
&&  \quad  
\times 
    \Bigl( 36  - (\alpha +\delta )  (\alpha  (8 \delta  (\alpha +\delta )-19)+17 \delta )    \Bigr)  \Bigr)^{1/2}   \,,  \nonumber \\  
\eea \bea
&& B_1 =      (\alpha   +\delta ) \left(5 \alpha ^3+3 \alpha ^2 \delta -9 \alpha  \delta ^2+32
   \alpha -7 \delta ^3-16 \delta \right) \nonumber \\
&& \quad + 48    
+\Bigl(  [  4+ (3 \alpha -\delta ) (\alpha +\delta )] 
   \left[  (\alpha +\delta )^2+4\right]^2  \nonumber \\
&&  \quad  
     \times \bigl(  36-   (\alpha +\delta ) (\alpha  (8 \delta  (\alpha +\delta )-19)+17 \delta   )  \bigr) \Bigr)^{1/2} ,   \\ 
&& B_2 =   2 \Bigl(  2 \left(\alpha ^2+2\right) \delta ^2+\alpha 
   \left(3 \alpha ^2+4\right) \delta -4 \left(\alpha ^2+2\right) \nonumber \\
&&  \quad \quad  - \alpha  \delta^3 \Bigr)   \times \left((\alpha +\delta )^2+4\right) \,.
\eea
With this data, we can 
obtain the interpolating solutions which smoothly connect the IR hyperscaling violating geometries to 
the UV AdS$_4$ geometries. 
To connect our analytic IR hyperscaling violating solutions to an asymptotically AdS$_4$ metric, 
we numerically solve the Einstein equations  
from $r=r_0$ 
to $r=\infty$. 
Here $r_0$ is some small radius  
where the solutions are given by eq.~(\ref{deviation1}) and (\ref{deviation2}). 
We solve numerically by setting $V_0 = -1$.  
At $r = r_0$, the condition $|\epsilon d_i r_0^\nu|\ll 1~(i=1,2)$ needs to be satisfied  
\footnote{In the asymptotic AdS$_4$ region, 
$\phi$ behaves as $\phi\simeq D_+ r^{\lambda_+}+D_- r^{\lambda_-}$ at 
$r\to \infty$, where $\lambda_\pm = - 3/2 \pm \sqrt{9 - 24\delta^2 }/2$. 
As shown in \cite{Hertog:2004rz}, the theory is generally unstable for the perturbation with respect 
to $\phi$ when both modes exist. 
However for the perturbation analysis we conduct in this paper, 
the dilaton $\phi$ does not fluctuate. Therefore we consider that the instability of \cite{Hertog:2004rz}
is not related to the stripe instability we 
investigate in this paper. In addition, for the numerical analysis in \S \ref{section3D}, 
we choose $D_+$ as small as possible in the range 
$|\epsilon d_i r_0^\nu|\ll 1~(i=1,2)$.}. 
We can obtain the appropriate boundary condition in the UV by fine-tuning $\epsilon$. 

%

The results of interpolating functions $\tilde a(r)$, $b(r)$, $\phi(r)$,  
for the parameter choice $\alpha = -{1}/{3}$, $\delta = 0.55$, $\epsilon \, d_1 = 0.01$ are shown in Fig.~1 - 3. 
%
For this parameter choice, we obtain 
$\gamma \approx 1 - 0.06$, $\beta \approx 0.01$, $ k \approx -0.1$, 
$C_a \approx 1$, $Q_e^2 \approx 0.5$, and we have positive $\nu$, $\nu \approx 0.9 > 0$, and real $d_2$ and $d_3$, $d_2 \approx -1.4 \, d_1$, $d_3 \approx 0.5 \, d_1$.  
The results shown in Fig.~1 - 3 are  
essentially the same as given in the appendix F of \cite{Iizuka:2011hg}. 

\begin{figure}
 \begin{center}
 \includegraphics[scale=.85]{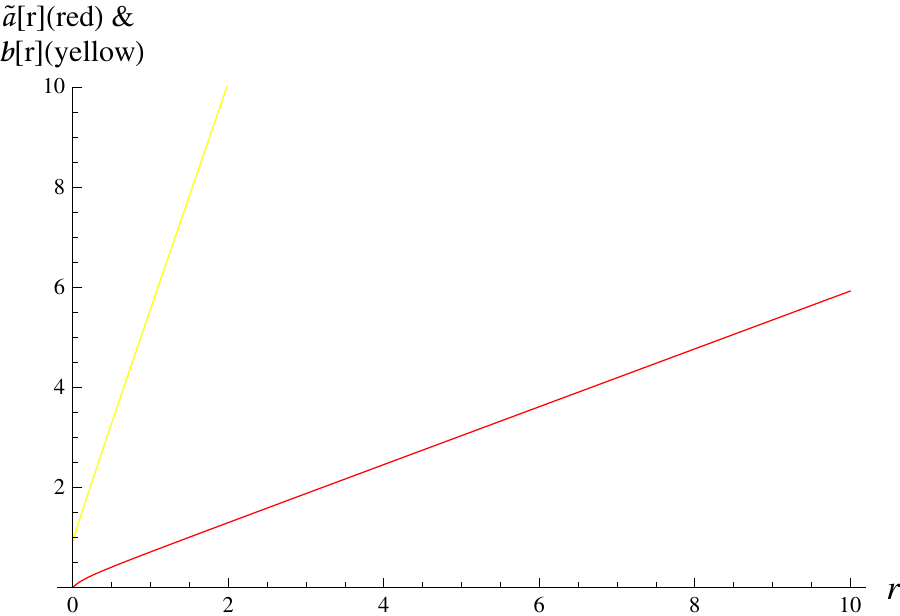}
  \caption{The interpolating solution $\sqrt{-g_{tt}(r)} = \sqrt{g^{rr}(r)}= \tilde a(r)$ (red) and $\sqrt{g_{xx}(r)} = \sqrt{g_{yy}(r)} = b(r)$ (yellow) for the 
 parameter choice $\alpha = -{1}/{3}$, $\delta = 0.55$, $d_1 = 0.01$, $V_0 = -1$.  
 It shows $\lim_{r \to \infty}\tilde a(r) \to \frac{1}{\sqrt{3}} r$, and $\lim_{r \to \infty} b(r) \propto r$ 
  as is expected from asymptotic AdS$_4$ in the UV. (color online)}
 \end{center}
\end{figure}

\begin{figure}
 \begin{center}
 \includegraphics[scale=.85]{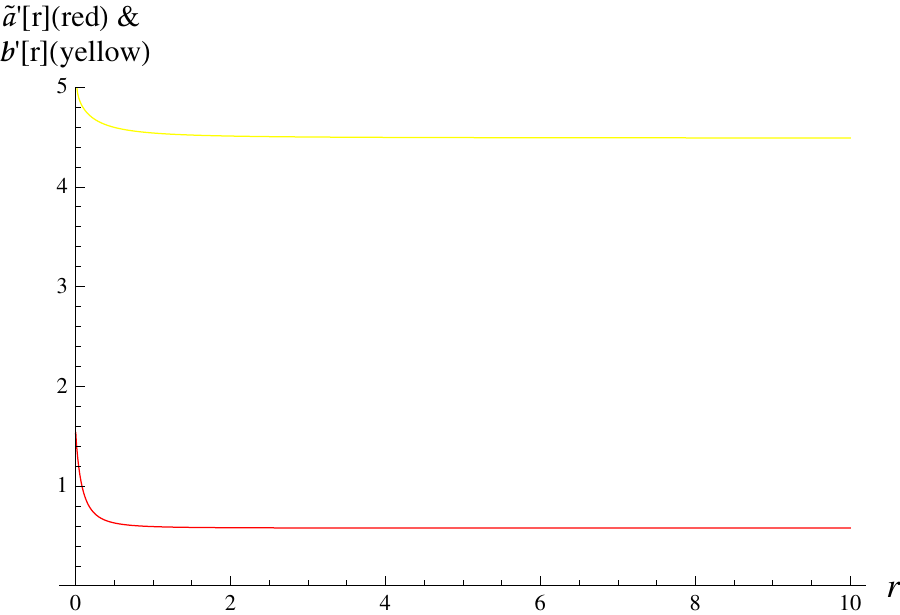}
  \caption{The interpolating solution $\tilde a'(r)$ (red) and $b'(r)$ (yellow) for the 
 same parameter choice. 
 $\lim_{r \to \infty} \tilde a'(r) \to \frac{1}{\sqrt{3}}$. 
 (color online)}
 \end{center}
\end{figure}

\begin{figure}
 \begin{center}
 \includegraphics[scale=.82]{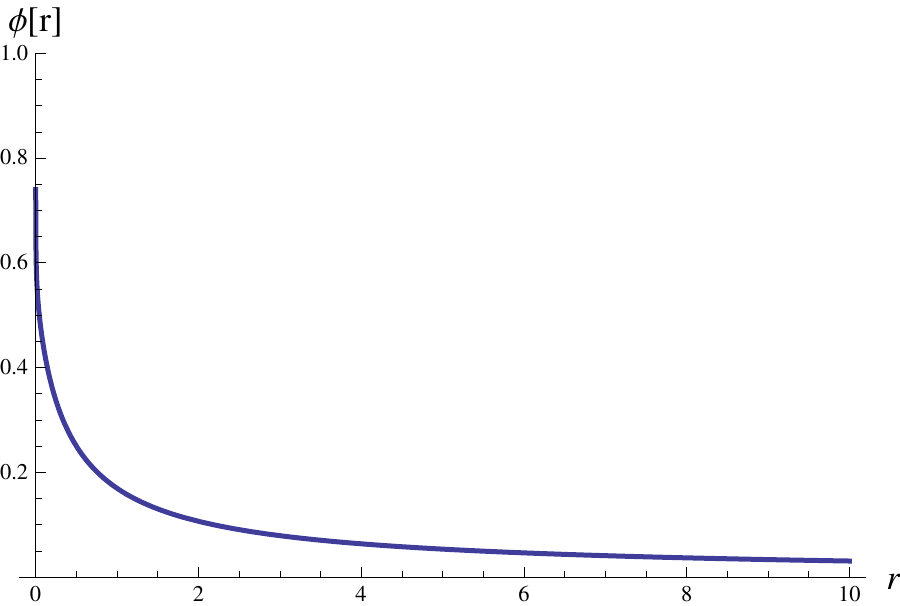}
  \caption{The interpolating solution $\phi(r)$ 
  for the 
 same parameter choice. $\lim_{r \to \infty} \phi(r) \to 0$, for the asymptotic AdS$_4$ in the UV. (color online)}
 \end{center}
\end{figure}

Similarly in the finite temperature case, we can obtain the interpolating solutions.  
We now add the $x$-dependent perturbation on this background 
to study the 
stripe stability on the background geometry. 

\section{Perturbation analysis in the probe limit}
\subsection{Perturbation equations in the probe limit}
\label{section3A}

Given the full solutions which interpolate between the IR hyperscaling violating geometries and UV AdS$_4$, 
we add the $x$-dependent perturbation and study the onset of the stripe instability. 
First, for the sake of simplicity, we neglect the graviton fluctuation,  
namely, we consider the stability in the ``probe limit'', where matter fluctuation will not 
back-react to the graviton fluctuation. 
In this limit, the translational symmetry breaking can occur 
and there can exist unstable mode,  
but the background geometries are fixed. 
We will see even in this case, we have interesting phenomena for a 
dynamical instability of the translationally invariant vacuum. 

On the background metric and flux given by eq.~(\ref{metricansatz}) and eq.~(\ref{fluxansatz}), we add the following small fluctuation, 
\bea
\delta A_y \quad \,, \quad \delta a \quad \,, \quad \delta \phi \,, 
\eea
and we choose these fluctuations are $x$-coordinate dependent. 
If any of these modes show nonzero vacuum expectation value with the boundary condition that their 
non-normalizable modes disappear at the boundary, we have gravity dual of the translational symmetry 
breaking.

From the dilaton fluctuation equation eq.~(\ref{dilatonEOM}), we have 
\bea
\frac{4}{\sqrt{-g}} \partial_\mu (\sqrt{-g} g^{\mu\nu} \partial_\nu \delta \phi)  = 
\partial_\phi ^2 V(\phi, a) \delta \phi 
\nonumber \\
+ (\partial_\phi^2 f(\phi) ) F^2 \delta \phi +  2 (\partial_\phi f(\phi) ) F^{\mu\nu} \delta F_{\mu\nu} \,. 
\eea
Note that the third term on the RHS vanish, since $F^{\mu\nu} \neq 0$ only for $F^{tr}$ 
but we excite only $\delta A_y$. 
Therefore, dilaton equation of motion can be trivially satisfied by setting 
\bea
\delta \phi = 0 \,, 
\eea
so we can set the dilaton fluctuation to be zero. 
In this paper, we will set the fluctuation of $\phi$ to be zero, $\delta \phi = 0$. 
In such case, the operators which are dual to $A_y$, $a$ can condensate.

The axion fluctuation equation of motion eq.~(\ref{axionEOM}) gives
\bea
\label{axionfluc}
\frac{4}{\sqrt{-g}}  \partial_\mu (e^{2 \xi \phi}\sqrt{-g} g^{\mu\nu} \partial_\nu \delta a )  
\quad \quad && \nonumber \\
 \quad =
( \partial_a \theta(a) )  \epsilon^{\mu\nu\lambda\eta} F_{\mu\nu} \delta F_{\lambda \eta} \,.  &&
\eea
Here we have used  
the fact that background flux is purely electrical, $F \tilde F = 0$. 

The gauge field equation of motion eq.~(\ref{EOMforgaugeboson}) yields, 
\bea
\label{gaugebosonfluc}
&& \partial_\mu \Bigl( \sqrt{-g} \bigl( f(\phi) \delta F^{\mu\nu} + \theta(a) \delta \tilde F^{\mu\nu} \nonumber\\
&& \quad \quad \quad  \quad  + (\partial_a \theta(a) ) \tilde F^{\mu\nu} \delta a \bigr)  \Bigr)  = 0 \,, 
\eea
here we used  
 $\delta \phi = 0$. 
We need to study the coupled equation for $\delta A_y$ and $ \delta a$ given by eq.~(\ref{axionfluc}) and eq.~(\ref{gaugebosonfluc}).

We follow the analysis of \cite{Donosstripes} closely. We set, 
\bea
\label{nogravitonflucAya}
\delta A_y &=& \delta A_y(r) \sin {q x} \, e^{- i \omega t }\,, \\
\label{nogravitonfluca}
\delta a &=& \delta a(r) \cos {q x} \, e^{- i \omega t} \,.
\eea

By using the background metric and flux given previously, after a bit algebra, 
the $\nu = y$ component of eq.~(\ref{gaugebosonfluc}) gives,   
\bea
&&\sqrt{-g} g^{yy} f(\phi) \left( - g^{xx} q^2 - g^{tt} \omega^2 \right) \delta A_y(r) \nonumber \\
&& \quad + \partial_r \left( \sqrt{-g} g^{yy} f(\phi)  g^{rr} \partial_r  \delta A_y(r) \right) \nonumber \\
&& \quad \quad -q   \sqrt{-g}
(\partial_a \theta(a)) \tilde F^{xy} \delta a(r) = 0 \,.
\eea
With the metric ansatz eq.~(\ref{metricansatz}), and by 
 using $\delta \tilde F^{\mu y} = 0$, and   
\bea
\sqrt{-g} \tilde F^{xy} = \sqrt{-g} \epsilon^{xytr} F_{tr}  
= - \frac{Q_e}{b^2 f(\phi)} \quad \,, 
\eea
this equation is re-written as   
\bea
\label{coupledgaugeaxionone}
&& f(\phi) \left( - b^{-2} q^2 + \tilde a^{-2} \omega^2  \right) \delta A_y(r)  \nonumber \\
&& \quad + \partial_r \left(f(\phi) \tilde a^2 \partial_r  \delta A_y(r) \right) 
\nonumber \\
&& \quad \quad + 
(\partial_a \theta(a)) \frac{q Q_e}{b^2 f(\phi)}   \delta a(r) = 0 \,.
\eea
Note that both a term proportional to $q^2$, and a term induced by axion term, have explicit 
common factor $b^{-2}$.  

Using $\epsilon^{rtxy}  = - \epsilon^{xtry}  = 1/\sqrt{-g}$ and the fact that the background axion takes
 the zero VEV, 
it is straightforward to check that $\nu \neq y$ component of eq.~(\ref{gaugebosonfluc}) is automatically satisfied. 

Similarly, the axion fluctuation equation of motion is written as 
\bea
\label{coupledgaugeaxiontwo}
&&  e^{2 \xi \phi} b^2 \left( -  b^{-2} q^2 + \tilde a^{-2} \omega^2 \right) \delta a(r)  \nonumber  \\
&& \quad + \, \partial_r (( e^{2 \xi \phi} b^2 ) \tilde a^2 \partial_r \delta a(r) )  \nonumber \\
&& \quad \quad +
 ( \partial_a \theta(a) ) \frac{q Q_e}{b^2 f(\phi)}\delta A_{y}(r)  = 0 \,.  
 \eea

We will solve the coupled equations eq.~(\ref{coupledgaugeaxionone}) and eq.~(\ref{coupledgaugeaxiontwo}) 
to study the translational symmetry breaking and stability of the translationally invariant vacuum. 
However since eq.~(\ref{coupledgaugeaxionone}) and  
(\ref{coupledgaugeaxiontwo}) are coupled, 
generically it is quite difficult to solve these for the stability. 
However since there are analytical expressions for the metrics in the IR,   
which are the hyperscaling violating geometries given by eq.~(\ref{finiteTtildear}) and (\ref{finiteTb}), 
we first restrict our attention to the near horizon IR geometries. 
 
\subsection{Negative momentum square mode on the hyperscaling violating geometry: 
finite temperature analysis} 
\label{section3B}


First, let's consider the finite temperature case, 
where the near horizon metric is 
given by  eq.~(\ref{finiteTtildear}) and (\ref{finiteTb}). 
In the near horizon region, where $r \to r_h$, we can approximate the metric as 
\bea
&& \tilde a^2 = (2 \beta + 2 \gamma -1 ) C_a^2 r_h^{2 \gamma-1} \left( r - r_h  \right)   \,, \\
&& b^2 \to r_h^{2 \beta} \,, \quad f(\phi) \to r_h^{2 \alpha k} \,, \quad  
\label{thebehaviorofb2andfphi}
\eea
With this approximation, the eq.~(\ref{coupledgaugeaxionone}) and eq.~(\ref{coupledgaugeaxiontwo}) reduce to 
\bea
&& r_h^{2 \alpha k} \left( - r_h^{-2 \beta} q^2 
+  \tilde a^{-2} \omega^2  \right) \delta A_y(r)  \\
&& \quad + \partial_r \left(r_h^{2 \alpha k} \tilde a^2 \partial_r  \delta A_y(r) \right) 
+ c_1 \frac{q Q_e}{r_h^{2 \beta + 2 \alpha k}}   \delta a(r) = 0 \,. \nonumber \\
&&    r_h^{2 \xi k + 2 \beta} \left(   - r_h^{-2 \beta} q^2 + \tilde a^{-2} \omega^2\right) \delta a(r)    \\
&&  \quad + \partial_r (r_h^{2 \xi k + 2 \beta}  \tilde a^2  \partial_r \delta a(r) )  
+
c_1 \frac{q Q_e}{r_h^{2 \beta+ 2 \alpha k}}\delta A_{y}(r)  = 0 \,.  \nonumber 
\eea

It is more convenient to re-write these as 
\bea
&& \left( - r_h^{-2 \beta} q^2 +  \tilde a^{-2} \omega^2  \right) \delta A_y 
 + \partial_r \left(  \tilde a^2 \partial_r  \delta A_y \right) \quad \quad  \nonumber \\
 && \quad \quad \quad + c_1 \frac{q Q_e}{r_h^{2 \beta + 4 \alpha k}}   \delta a = 0 \,,   \\
&& \left( - r_h^{-2 \beta} q^2 +  \tilde a^{-2} \omega^2  \right) \delta a
 + \partial_r \left(  \tilde a^2 \partial_r  \delta a \right) \quad \quad  \nonumber \\
&& \label{secondeqforc2multi}  \quad \quad \quad +  c_1 \frac{q Q_e}{r_h^{4 \beta + 2 \xi k +2 \alpha k}}   \delta A_y = 0 \,.  
\eea
By multiplying 
$c_2 = r_h^{\beta + \xi k - \alpha k}$ to eq.~(\ref{secondeqforc2multi}), 
we obtain 
\bea
 \left( - r_h^{-2 \beta} q^2 +  \tilde a^{-2} \omega^2  \right) \psi_1
 + \partial_r \left(  \tilde a^2 \partial_r  \psi_1 \right) \quad \quad \quad \quad &&  \nonumber \\
 \quad + c_1 \frac{q Q_e}{c_2 r_h^{2 \beta + 4 \alpha k}}   \psi_2 = 0 \,, && \\
 \left( - r_h^{-2 \beta} q^2 +  \tilde a^{-2} \omega^2  \right) \psi_2
 + \partial_r \left(  \tilde a^2 \partial_r  \psi_2 \right) \quad \quad \quad \quad && \nonumber \\
 \quad + c_1 \frac{q Q_e}{c_2 r_h^{2 \beta + 4 \alpha k}}   \psi_1 = 0 \,. && 
\eea
where
\bea
\label{psionepsitwodefinition}
\delta A_y \equiv \psi_1 \quad \,, \quad c_2 \, \delta a \equiv \psi_2
\eea
By choosing 
$\psi_{\pm} \equiv \psi_1 \pm \psi_2$, 
we have two independent equations 
\bea
\left( - r_h^{-2 \beta} q^2 +  \tilde a^{-2} \omega^2  \right) \psi_{\pm} 
 + \partial_r \left(  \tilde a^2 \partial_r  \psi_{\pm} \right) \quad \quad \quad \quad \nonumber \\ \pm 
c_1 \frac{q Q_e}{r_h^{3 \beta + 3 \alpha k + \xi k}}   \psi_{\pm} = 0 \,. 
\eea
These equations have very similar structure to the equations of motion for massive scalars in the 
finite temperature black brane. However due to the axion term proportional to $c_1$, 
it has an 
``effective momentum square'' defined as 
\bea
\label{negativemasssquare}
q^2_{eff \pm} &\equiv&  
q^2 \mp c_1 \frac{q Q_e}{r_h^{  \beta + 3 \alpha k + \xi k}} \,.
\eea
We take $c_1 q Q_e > 0$ without loss of generality. 
Note that this guarantees that at the horizon, one of the ``effective momentum square'', 
$q^2_{eff +} $ above for the mode $\psi_+(r)$,   
becomes always negative.   
The minimum value is given at critical wave number $q_c$, 
\bea
q_c = \frac{c_1 Q_e}{2 r_h^{  \beta + 3 \alpha k + \xi k}} \,, 
\eea
with minimum momentum square 
\bea
\label{theminimalmomentumsqvalue}
q^2_{eff +} |_{q = q_c}
=   - 
\frac{ c^2_1Q^2_e}{4 r_h^{ 2( \beta 
+ 3 \alpha k + \xi k)}} < 0 \,,
\eea
which is always negative. 

This is very similar to the situations where the striped phase instability 
is studied by \cite{ooguri, Donosstripes}; 
The axion term $c_1 \neq 0$ makes the system have lower energy by having the 
nonzero momentum ``stripe'' type of the spatially modulated mode. 

However contrary to the IR AdS$_2$ case, 
there are difference on the study of the instability of the IR hyperscaling violating geometries   
with finite but small temperature;  
Due to the additional warping factor in hyperscaling violating geometries, 
there is extra temperature ($r_h$) dependence for this effective momentum square, and 
$c_1 Q_e$  appears only in the combinations of $c_1 Q_e/{ r_h^{ \beta 
+ 3 \alpha k + \xi k }}$ \footnote{$Q_e$ is actually not a parameter, since it takes definite value as eq.~(\ref{case2}).}. In the AdS$_2$ case, where $\alpha = \delta = 0$ for $\beta = k = 0$ and $\gamma = 1$ in eq.~(\ref{case1}) and (\ref{case2}), 
$c_1 Q_e$ does not involve any temperature dependence, 
since ${ r_h^{ \beta + 3 \alpha k + \xi k }}$ becomes constant. 

If  $c_1 Q_e/{ r_h^{ \beta + 3 \alpha k + \xi k }}$ are very large, 
for $q \neq 0$, $\psi_+$ mode can have lower energy with nonzero momentum $q$ mode, 
and the existence of such negative momentum square mode 
gives the possibility that such mode induces the instability of the translationally invariant vacuum.

The fact that $c_1 Q_e$  appears in the combinations of $c_1 Q_e/{ r_h^{ \beta 
+ 3 \alpha k + \xi k }}$ implies that 
the situation is very different, depending on $ \beta + 3 \alpha k + \xi k$ is positive or negative. 
If 
\bea
\label{finiteTcriteriaforeffectivemomentumenhancement}
\beta + 3 \alpha k + \xi k > 0 \,,
\eea 
as we lower the temperature ({\it i.e.}, as we lower $r_h$), the instability are more likely 
to occur even for the small values of $c_1 Q_e$, due to the enhancement by the factor 
$1/{ r_h^{ \beta + 3 \alpha k + \xi k }}$. 
In this case, 
$c_1 Q_e/{ r_h^{ \beta + 3 \alpha k + \xi k }}$ diverges positively at $r_h \to 0$ and $q^2_{eff +}$ 
goes to negative infinity, and 
these suggest that the instabilities are expected to occur.   
On the other hand, if $ \beta + 3 \alpha k + \xi k$ is negative, as we lower the temperature, we need to 
increase $c_1 Q_e$ in order to 
keep the same value for $q^2_{eff +}$, due to the suppression by the factor $1/{ r_h^{ \beta + 3 \alpha k + \xi k }}$. 

However, 
whether the existence of the negative effective momentum square mode, $q^2_{eff +} < 0$, 
which is evaluated at the horizon, is enough or not for instability is another question. 
This is because in general hyperscaling violating geometries, we do not know 
the critical values of the instability. This is in sharp contrast to the AdS$_2$ case. 
In the AdS$_2$ case, partially due to the scale invariance of the geometry, we have 
a sharp bound for the instability, {\it i.e.,} Breitenlohner - Freedman (BF) bound.  However for geometries with hyperscaling violation, 
we generically do not expect sharp bound for generic radius and 
therefore in order to show that there is an unstable mode, local argument is not enough in general 
and we need to study the full bulk system to search for the unstable mode.  
Note that since hyperscaling violating geometries are IR limit and they approach AdS$_4$ in the UV, 
this means that we need to study the perturbation equations on the whole geometries, 
with the normalizable boundary condition in the UV AdS$_4$ region. 

However, before we study the full bulk system for the unstable mode numerically,  
one might wonder if something special occurs for the special parameter range where the equality of 
eq.~(\ref{finiteTcriteriaforeffectivemomentumenhancement}) holds. 
To understand this marginal parameter range in more detail, we continue 
the study of perturbation equations on the IR hyperscaling violating geometries by taking the zero temperature limit.

\subsection{Zero temperature 
analysis and instability criteria}
\subsubsection{Zero temperature perturbation} 
\label{section3C1}

We analyze the instabilities at zero temperature limit, where the metric is 
\bea
\tilde a^2 = C_a^2 r^{2 \gamma} \quad \,, \quad b^2 = r^{2 \beta} \,.
\eea
In this case, by redefining the field as 
\bea
\delta A_y \equiv r^{1 - \gamma - \alpha k} { {\delta \hat A_y}} \quad \,, \quad 
\delta a \equiv r^{1 - \beta - \gamma - k \xi} \delta \hat a \,,
\eea
the gauge field and axion equations of motion are written as 
\bea
\label{M2M2M2M2}
\nabla^2 \delta \hat X = M^2 \delta \hat X
\eea
where
\bea
\nabla^2  &\equiv& C_a^2 \partial_r   r^2 \partial_r   \,, \quad \\
\label{thedefinitionofmatrixm2forABCD}
\delta \hat X  &=& \left(   \begin{array}{cc}  { {\delta \hat A_y}}  \\ 
 { {\delta \hat a}}  \end{array} \right) \,, \, 
M^2 = \left( \begin{array}{cc} 
A(r) & B(r)  \\
B(r) & C(r)   
\end{array} \right) \,, \\
A(r) &=&    
C_a^2 \left(-1 + \gamma + \alpha k \right) \left(\gamma + \alpha k \right)
\nonumber \\ && \quad  + 
  q^2 r^{2 -2  \beta - 2\gamma}  \,, \\
B(r) &=&  
 - c_1 q Q_e r^{2 - 3 \beta - 2 \gamma - k (3 \alpha + \xi)}  \,, \\
C(r) &=& C_a^2 \left( -1 + \beta + \gamma + k \xi \right) \left( \beta + \gamma + k \xi \right) \nonumber \\ 
&&  \quad +   q^2 r^{2 -2 \beta -2 \gamma}  \,,  
\label{thedefinitionofmatrixm2forD}
\eea
where we have set $\omega = 0$, since it allows us to see the instability onset. 

Note that  if both 
\bea
\label{firstcondition}
2 - 3 \beta - 2 \gamma - k (3 \alpha + \xi) < 0  \,,
\eea
and 
\bea
\label{secondcondition}
2 - 3 \beta - 2 \gamma - k (3 \alpha + \xi) < 2 -2 \beta -2 \gamma \,,
\eea
are satisfied, 
the off-diagonal terms, which are proportional to $c_1$ and which come from the axion term, 
dominate at the extremal horizon, $r \to 0$. In such case, 
we are forced to have the situation where the matrix $M^2$ has at least one negative eigenvalue, 
which goes to negative infinity at $r \to 0$.   
Therefore in this parameter range, it is expected that there is an unstable mode for the 
stripe instability.  

On the other hand, if the parameters of hyperscaling violation do not satisfy above inequality 
(\ref{firstcondition}) and (\ref{secondcondition}), then the matrix $M^2$ does not always have 
negative eigenvalue at $r \to 0$. However even in this case, if we take large $c_1$, then 
at finite $r$, $M^2$ has one eigenvalue which becomes negative at some intermediate radius $r$. 
From these, we expect that if inequality 
(\ref{firstcondition}) and (\ref{secondcondition}) hold, the instability is more likely to occur compared 
to the case where nequality 
(\ref{firstcondition}) and (\ref{secondcondition}) do not hold. We will see this actually later in the numerical analysis. 
 
Note that the second condition eq.~(\ref{secondcondition}) is equivalent to 
\bea
\label{thisisthesecondone}
\beta + 3 \alpha k + \xi k > 0 \,,
\eea
which is the same as eq.~(\ref{finiteTcriteriaforeffectivemomentumenhancement}), {\it i.e.,}
the second term of effective negative momentum square in eq.~(\ref{negativemasssquare}), 
dominates at the extremal limit $r_h \to 0$.   
The first condition eq.~(\ref{firstcondition}) is to guarantee that the off-diagonal term will not vanish at the horizon. 

\subsubsection{Analytical criteria for instability onset for special type of hyperscaling violation}
\label{section3C2}

There is a special parameter range in hyperscaling violating geometries, which occurs
when the equality of both eq.~(\ref{firstcondition}) and (\ref{secondcondition}) are satisfied. 
This is when  $\xi$, the parameter for the axion kinetic term in eq.~(\ref{actionofourmodel}), is tuned as 
\bea
\label{xiistuned}
 \xi  = - 3 \alpha - \frac{ 3 \beta  + 2 \gamma - 2}{k} \,, 
\eea
and 
$\beta$ and $\gamma$ in the IR hyperscaling violating geometries satisfy  the relation 
\bea
\label{betagammaequalone}
\beta+ \gamma = 1 \,. \quad
\eea
For this special type of hyperscaling violating geometries, 
we can identify the critical mass just as Breitenlohner - Freedman (BF) bound, 
since the matrix $M^2$, given in eq.~(\ref{thedefinitionofmatrixm2forABCD}), becomes independent on the radius $r$, and therefore constant matrix.  
From the solution eq.~(\ref{case1}) and (\ref{case2}), 
we can see that if we do not have parameter $\xi$, the only way to satisfy eq.~(\ref{xiistuned}) and (\ref{betagammaequalone}) are AdS$_2$ case, {\it i.e.,} $\beta= 0$ and $\gamma = 1$ case. 
However for the theories with tuned axion kinetic terms $\xi$ as eq.~(\ref{xiistuned}), $M^2$ becomes 
constant matrix for one real parameter family of hyperscaling violating geometries with eq.~(\ref{betagammaequalone}), and AdS$_2$ is a special parameter point in this parameter range.

For such IR geometries with hyperscaling violation satisfying (\ref{betagammaequalone}), 
$M^2$ has two eigenvalues,  $m^2_{\pm}$, 
\bea
\label{theM2eigenvalueplusminusnotation}
m^2_{\pm}(q) =  \frac{1}{2} \left( A_0 + C_0 \pm \sqrt{(A_0 - C_0)^2 + 4 B_0^2}  \right) \,,
\eea
where 
\bea
\label{thisisA0foranaly}
A_0 &=&    
C_a^2 \left(-1 + \gamma + \alpha k \right) \left(\gamma + \alpha k \right)
 +   q^2   \,, \\
B_0 &=&   - c_1 q Q_e   \,,  \\
C_0 &=& C_a^2 \left( -1 + \beta + \gamma + k \xi \right) \left( \beta + \gamma + k \xi \right) +   q^2  \,. 
\label{thisisC0foranaly}
\eea
$\delta \hat X$ admits the power law ansatz 
\bea
\delta \hat X \sim r^\Delta \,,
\eea 
where $\Delta$ satisfies 
\bea
C_a^2 \Delta (\Delta + 1) =  
m^2_{\pm}(q) \,,
\eea
and 
critical values for the instability onset are given when $\Delta$ becomes 
complex \footnote{This simple method gives the generalized 
Breitenlohner - Freedman (BF) bound 
for the stability in non-AdS case, for examples, \cite{unpublishedwork, Gath:2012pg}.}. 
By minimizing $m_{-}^2(q)$ with respect to 
$q$ at $q = q_{min}$, it becomes 
\bea
C_a^2 \Delta (\Delta + 1) = m^2_{-}(q_{min}) \,,
\eea
where $q_{min}$ satisfies $\partial_q m_{-}^2(q) |_{q = q_{min}}=0$. 
$\Delta$ becomes complex when 
\bea
\label{complexconditionforanalitical}
{m_{-}^2}(q_{min}) \le - \frac{1}{4} {C_a^2}  
\eea
holds. 
In the large $c_1Q_e$ and $q$ case, 
\bea
m^2_{-}(q) \approx
 q^2 -  |c_1 Q_e| q  \,,
\eea
which allows $q_{min} \approx |c_1 Q_e| /2 $, and $m^2_{-}(q_{min}) \approx - |c_1 Q_e|^2 /4$. Therefore 
in this case, eq.~(\ref{complexconditionforanalitical}) is satisfied and we have stripe instability. 

The special relation eq.~(\ref{betagammaequalone}) is  
the point where both $z$ and $\theta$ diverge ($z$ and $\theta$ 
are given by eq.~(\ref{thedefinitionofzandthetaforhyper})). More precisely, 
it is the special point of hyperscaling violation where either $z \to - \infty$ and $\theta \to + \infty$, or 
$z \to + \infty$ and $\theta \to - \infty$ holds. 

It is interesting to see how above results are different from the AdS$_2$ case.  
Let's first consider the AdS$_2$ case, which is the special limit of above and  
we take $\alpha = \delta = 0$ so that we have $\beta = 0$, $\gamma = 1$, $k = 0$ in eq.~(\ref{case1}), and eq.~(\ref{case2}). 
In such case, $A_0$, $B_0$, $C_0$ in eq.~(\ref{thisisA0foranaly}) - (\ref{thisisC0foranaly}) simplifies 
and we obtain 
\bea
m^2_{\pm AdS2}(q) = q^2 \pm  c_1 q Q_e \,.
\eea
This $m_{\pm AdS2}^2$ is the same effective momentum square $q^2_{eff \mp}$ given in eq.~(\ref{negativemasssquare}).
Therefore in AdS$_2$ case, the onset value of the instability, which we call $c_{min \, AdS_2}$,  
is given by the equality of eq.~(\ref{complexconditionforanalitical}) as 
\bea
\label{cminads2case}
c_{min \, AdS_2} = \frac{C_a|_{\alpha = \delta = 0} }{Q_e|_{\alpha = \delta = 0} } = \sqrt{2} \,.
\eea
Here we have used eq.~(\ref{case2}) and (\ref{case3}). 
The stripe instability occurs at $c_1 \ge c_{min \, AdS_2}$. 

Next, we consider non-AdS$_2$ case. 
Note that eq.~(\ref{betagammaequalone}) is satisfied when $\alpha  = \pm \delta$ 
from eq.~(\ref{case1}) and (\ref{case2}). However since $\alpha = - \delta$ gives $\beta = 0$, $\gamma = 1$, 
which is AdS$_2$, we consider $\alpha = \delta$ case.  From eq.~(\ref{case1}) - (\ref{case3}), in this case 
the hyperscaling violating geometries are parametrized by only one real parameter $\alpha$, with 
\bea
&&\beta = \frac{\alpha^2}{1 + \alpha^2}   \,, \quad \gamma = \frac{1}{1 + \alpha^2}  \,, \quad 
k = - \frac{\alpha}{1 + \alpha^2} \,, \quad \\ 
&& Q_e^2 = - V_0 \frac{1 - \alpha^2}{2 (1 + \alpha^2)}   \,, \quad C_a^2 = - V_0   \,, 
\eea
with $V_0 < 0$, and eq.~(\ref{xiistuned}) gives 
\bea
\xi = - 2 \alpha \,. \quad
\eea
Then, the eigenvalue $m^2_-(q)$ of $M^2$ given by eq.~(\ref{theM2eigenvalueplusminusnotation}) becomes 
\bea
 m^2_-(q) &=&  \frac{1}{2 \left(\alpha ^2+1\right)^4} \Bigl(
 2 \left(\alpha ^2+1\right)^4 q^2 
 - 8\left(\alpha ^2+1\right)^2 \alpha ^4 V_0 
 \nonumber \\
&& - \sqrt{2} \sqrt{\left(\alpha ^2+1\right)^6 V_0
   \left(\left(\alpha ^4-1\right) c_1^2 q^2+8 \alpha ^4 V_0\right)} 
\Bigr) \,. \nonumber \\
\eea
Since this has minima at 
\bea
q = q_{min} = \frac{\sqrt{- V_0} \sqrt{\left(1 - \alpha ^2\right)^2 c_1^2 - \frac{64 \alpha ^4}{c_1^2}}}{2
    \sqrt{2 (1 - \alpha ^4) } } \,,
\eea
the instability condition eq.~(\ref{complexconditionforanalitical}) becomes 
\bea
&&\frac{  -2 \sqrt{\left(\alpha ^2-1\right)^2 \left(\alpha ^2+1\right)^6
   c_1^4 V_0^2}}{8 \left(\alpha ^2+1\right)^4} \nonumber \\
&& + 
\frac{\left(\alpha ^2+1\right)^2 V_0}{ 8  c_1^2  \left(\alpha ^2-1\right)
  \left(\alpha ^2+1\right)^4}  \,
 \times  \Bigl(\left(\alpha ^2-1\right)^2
   \left(\alpha ^2+1\right) c_1^4 \nonumber \\ 
&&   -64 \left(\alpha ^6+\alpha ^4\right)-32
   \left(\alpha ^2-1\right) \alpha ^4 c_1^2 \Bigr) 
\le  \frac{V_0}{4} \,.
\eea
Therefore, the onset $c_1$ value of the instability, which we call $c_{min}^\alpha$, 
is given by the equality of this and 
it is 
\bea
c_{min}^\alpha = \sqrt{2} \left( \frac{ {1 + \alpha ^2}}{{1 - \alpha^2}} \right)^{1/2}  \,. 
\eea
The instability occurs when $c_1 \ge c_{min}^\alpha$, 
which is larger than the critical value $c_{min \, AdS_2}$ for AdS$_2$ case, $\sqrt{2}$.  
Therefore it is more stable than the AdS$_2$ case, regarding the stripe instability. 

This expression loses its meaning when $\alpha^2 \ge 1$, corresponding to $\gamma \le 1/2$.   
But $\gamma$ cannot be less than $1/2$, since then the entropy density of these black brane 
are proportional to the negative power of the temperature, which, thermodynamically, does not make 
sense 
\footnote{See eq.~(\ref{Thinregardstorh}), 
if $\gamma < 1/2$, then $r_h \to 0$ with $T \to \infty$.}.  
Another reason for $\gamma > 1/2$ is that if $\gamma \le 1/2$, then the null rays from the zero temperature horizon $r = 0$ can reach nonzero $r$ at finite time, which contradicts with the ``horizon'' property at $r = 0$  
\footnote{We thank Noriaki Ogawa for pointing this out to us.}. 

\subsubsection{$\xi = 0$ case} 
\label{section3C3}

In the next section, we conduct numerical analysis of generic parameter points 
where such an equality eq.~(\ref{betagammaequalone}) does not hold. 
For that purpose, let's investigate the $\xi = 0$ case a bit more, since this is the case where 
axion has canonical kinetic term.  

The parameter range which satisfies 
eq.~(\ref{secondcondition}) is given by 
\bea
\label{5alphaminusdelta}
\left( \alpha + \delta  \right) \left( 5 \alpha - \delta \right) < 0 \,,
\eea
and for eq.~(\ref{firstcondition}), it is given by 
\bea
\label{3alphaplusdelta}
\left( \alpha + \delta  \right) \left( 3 \alpha + \delta \right) < 0 \,.
\eea
This gives 
\bea
\label{alphapositivecase}
-3 \alpha < \delta < -\alpha \quad (\mbox{for} \, \alpha > 0) \,, \\
-3 \alpha > \delta > -\alpha \quad (\mbox{for} \, \alpha < 0) \,.  
\label{alphanegativecase}
\eea

This range is written in terms of $\beta$ and $\gamma$, or $\theta$ and $z$. 
From eq.~(\ref{case1}), we have 
%
\bea
\label{thisistheplussigncase}
\alpha = \pm \frac{1 - 2 \beta - \gamma}{\sqrt{\beta (1 - \beta)}} \,, \quad 
\delta= \pm \frac{ \gamma -  1 }{\sqrt{\beta (1 - \beta)}} \,.
\eea
Remember that we need 
\bea
\frac{1}{2} < \gamma \,, \quad 0 < \beta < 1 \,. 
\eea 
Therefore, the parameter range, eq.~(\ref{alphapositivecase}) or (\ref{alphanegativecase}), is equivalent to 
\bea
\label{finalbetagammarangeforenhanceinstability}
0 < \beta < \frac{1}{3} (1 - \gamma)  \,,  \quad \frac{1}{2} < \gamma < 1 \,.
\eea
%
In terms of $\theta$ and $z$, 
using 
\bea
\beta = \frac{\theta - 4}{2 (\theta - 2 z)}  \,, \quad \gamma = \frac{\theta - 4 z}{2 (\theta - 2 z)} \,, 
\eea
from eq.~(\ref{thedefinitionofzandthetaforhyper}), the parameter range eq.~(\ref{finalbetagammarangeforenhanceinstability}) is written as 
\bea
4 < \theta < 6 \,, \quad z < 0 \,.
\eea

So far we consider the parameter range satisfying eq.~(\ref{firstcondition}) and (\ref{secondcondition}). 
In addition to these, we have more restriction on the parameters. 
First, we need $\delta \, k < 0$ which restricts $\delta ( \alpha + \delta) > 0$.   
Second, we want to study the 
stripe instability on the background which is stable at $q=0$. 
Since the IR hyperscaling violating geometries asymptotically approach AdS$_4$ in the UV, 
the mass of the dilaton must be above the AdS$_4$ BF bound for the stability at zero momentum $q = 0$. 
From our Lagrangian eq.~(\ref{actionofourmodel}), we can read off the mass of the dilaton at 
$\phi = 0$ and this gives additional constraint 
\bea
\label{deltasquare38forBF}
-\sqrt{\frac{3}{8} } < \delta < \sqrt{\frac{3}{8} }  \,.
\eea

Combined these additional conditions with 
eq.~(\ref{alphapositivecase}) and (\ref{alphanegativecase}), 
finally they are summarized as 
\bea
\label{alphadeltaconditionone}
0 > -\frac{1}{3} \delta > \alpha >  - \delta >  - \sqrt{\frac{3}{8} } \,. 
\eea
or, 
\bea
\label{alphadeltaconditiontwo}
0 < -\frac{1}{3} \delta < \alpha <  - \delta <  \sqrt{\frac{3}{8} }  \,.
\eea

It is clear that there are parameter ranges satisfying above 
in the ``physical parameter ranges'' \footnote{``Physical'' in the sense that $Q_e^2 > 0$, $C_a^2 > 0$, $\gamma > 0$, $\gamma - \beta > 0$, $\beta > 0$, and $\delta k < 0$ are satisfied.} parametrized by $\alpha$ and $\delta$, as is 
shown in Figure 1 of \cite{Iizuka:2011hg}. 
In this parameter range, the effective momentum square in eq.~(\ref{theminimalmomentumsqvalue}) 
becomes large negative value at low temperature, 
and also the off-diagonal components of matrix $M^2$ in eq.~(\ref{thedefinitionofmatrixm2forABCD}) 
dominates. 
Therefore it is expected that the instability is more likely to occur at small value of $c_1$ in this case. 

We next conduct numerical analysis for the parameter choice where eq.~(\ref{alphadeltaconditionone}) is satisfied and also for another parameter choice 
where it is not satisfied. We call the case where eq.~(\ref{alphadeltaconditionone}) is satisfied as case I, and the case where eq.~(\ref{alphadeltaconditionone}) (or eq.~(\ref{thisisthesecondone}) in 
more generic case where $\xi \neq 0$)
is not satisfied as case II. 

The limit $\alpha \to 0$ and $\delta \to 0$ corresponds to the 
$AdS_2$ and its stripe instability is well studied in \cite{gauntlett, Donosstripes}.  
We will now study numerically the whole bulk system in these parameter range for the stripe instability next.

\subsection{Numerical analysis for the bulk zero mode on the full geometries}
\label{section3D}

Now we conduct numerical investigation of the coupled 
equations eq.~(\ref{coupledgaugeaxionone}) and eq.~(\ref{coupledgaugeaxiontwo}) 
to study the dynamical translational symmetry breaking. 
So far we have been concentrating on the perturbation equation analysis restricting our attention 
on the IR hyperscaling violating geometries. However, 
now we will conduct the numerical analysis on the full geometries, 
which approach geometries  with hyperscaling violation in the IR and AdS$_4$ in the UV.  

In order to find the onset of the instability, we will look for the zero mode, namely the 
solution of eq.~(\ref{coupledgaugeaxionone}) and eq.~(\ref{coupledgaugeaxiontwo})  with 
$\omega = 0$. If there is such a zero frequency mode for some value of $c_1$, then it 
 is the 
critical mode. By increasing $c_1$ above the critical value, the instability occurs. 
This is because if there is an unstable mode, $\mbox{Im}(\omega)>0$ which grows as time evolves,  
then there should also be zero mode solution $\omega = 0$ at the instability onset point. 

Let us emphasize again why searching the zero mode 
on the full background geometries is important: 
In the case of IR AdS$_2$, we have a sharp local criteria for the onset of the stripe instabilities, 
which is given by the condition $c_1 \ge c_{min \, AdS_2}$ with eq.~(\ref{cminads2case}). This is in contrast to our 
generic hyperscaling violating geometries in the IR; in two-parameter ($\beta$ and $\gamma$, or $\theta$ and $z$) hyperscaling violating IR geometries, 
we do not have sharp criteria for the stripe instability from the IR geometries.  

However even in the IR AdS$_2$ case, it is important to find the zero mode on the 
full geometries which approach 
AdS$_4$ in the UV. 
This is because generically 
the stability is not determined by the IR region only, but rather it is determined by the full geometries.
It is especially so 
if  two modes are coupled.  

To see this, let us consider the IR AdS$_2$ geometries which approach AdS$_4$ in the UV. 
For the IR AdS$_2$ region, the matrix $M^2$ in eq.~(\ref{M2M2M2M2}) becomes constant matrix and we can 
obtain two eigenvector modes (let us call these as mode A and B) 
made by some specific linear combination of mode $\delta A_y$ and $\delta a$. 
Suppose that the mode $A$ gives the lower eigenvalue of the matrix $M^2$ than the mode $B$, 
and furthermore that the eigenvalue of the mode $A$ is lower than 
the AdS$_2$ BF bound and the eigenvalue of the mode $B$ is higher than the AdS$_2$ BF bound. 
It is true that if we excite only the mode $A$, 
then we can lower the energy by the mode $A$ perturbation and this indicates instability. 
However due to the UV boundary condition and the mixing of the two modes, generically we cannot excite only the mode $A$, but rather 
we need to excite the mode $B$ too in general \footnote{To see this, note that in the UV, we can 
similarly construct the matrix $M^2$ from the equations of motion for $\delta A_y$ and $\delta a$. 
The eigenvector mode in the UV 
(let us call these as $A'$ and $B'$)  are different from the eigenvector mode $A$ and $B$ (defined in 
IR) generically since the metric is different in UV and IR. 
The mode $A'$ and $B'$ are specific linear combinations of the mode $A$ and $B$.  
Now we need to impose boundary condition 
both at the IR horizon and UV boundary. At the horizon, we have ingoing boundary conditions 
or regularity condition for both $A$ and $B$, 
and at the UV boundary we impose that non-normalizable mode vanishes for both 
$A'$ and $B'$.  
Since for the second order differential equations, we have imposed two boundary conditions 
(one at the horizon and one at the boundary), we do not have parameter left and 
this generically implies that we are forced to have excitation both mode $A$ and $B$ in the IR near horizon region 
generically.}.   
At what ratio, we excite the mode $A$ and $B$, is determined by the full (whole) geometries.  
Therefore, depending on the ratio of the mode $A$ and $B$ excitations, we can see if the 
system is unstable or not by the perturbation. This is because the mode $A$ lowers the energy 
but the mode $B$ increases the energy in the IR AdS$_2$. In this way, it is generically determined not by the local IR geometry only, but rather by the full geometries. Therefore it is important to find the zero mode on the whole geometries including IR and UV. 

Furthermore, in the case of IR hyperscaling violating geometries, which is parametrized by two parameters, we do not generically have sharp BF-like bound. Therefore it is more important to search for the zero mode on the full geometries which include both IR and UV region. 

We now seek for the zero mode on the full geometries. 
For numerics, we set $V_0 = -1$. 
Introducing new variables $\psi_\pm=\delta A_y\pm c_2\psi_2$ as eq.~(\ref{psionepsitwodefinition}), 
we can obtain the boundary condition at the horizon $r=r_h$ by imposing regularity as  
\begin{align}
\label{bc_horizon}
\p_r\psi_\pm |_{r = r_h} 
= \frac{r_h^{1 - 2 \beta - 2 \gamma}q^2_{eff \pm} \psi_\pm|_{r = r_h}}
{(2\beta+2\gamma-1)C_a^2 }, 
\end{align} 
where $q^2_{eff \pm}$ is the effective momentum square defined in eq.~(\ref{negativemasssquare}). 


In order to find the onset of the spontaneous translational symmetry breaking in the holographic 
dual setting, we impose Dirichlet boundary condition for 
the variables $\delta A_y$ and $\delta a$, at the AdS$_4$ boundary, 
\begin{align}
\label{bc_infinity}
\delta A_y=\delta a=0 \qquad \mbox{for} \qquad r=\infty \,.
\end{align} 
This corresponds to the requirement that there is no non-normalizable mode for 
$\delta A_y$ and $\delta a$ in the UV AdS$_4$ boundary. Non-normalizable mode 
for $\delta A_y$ and $\delta a$ approach constant in the UV AdS$_4$.  

It is useful to consider the parameter counting. 
For given fixed parameters $\alpha$, $\delta$, $\xi$, $r_h$, and $c_1$, 
there are two free parameters $\psi_{-}|_{r = r_h}$ 
and $q$. 
Note that $\psi_{+}|_{r =r_h}$ is not a free parameter, since 
we can fix $\psi_{+}|_{r=r_h}$ to unity without loss of generality in linearized perturbations. 
We tune $\psi_{-}|_{r = r_h}$  
and $q$ such that 
the two boundary conditions eq.~(\ref{bc_infinity}) are satisfied. 
After this tuning, we have no parameter left, 
and as a results, we have nonzero normalizable mode for both $\psi_{+}$ and $\psi_{-}$ at $r \to \infty$. 
This implies that there are spatially modulated VEV for the scalar and vector current $<j_y>$, 
which are dual to axion and  and gauge boson $A_y$, and this implies that 
dual theories at IR show the ``current density wave'' phase. 
Note also that for a given temperature $T(r_h)$, we expect normalizable zero modes to appear 
at specific values of momentum $q$.

As we have discussed, we expect that the effect of the axion term and the negative ``effective momentum 
square'' is enhanced or suppressed depending on eq.~(\ref{alphadeltaconditionone}) 
is satisfied or not. 
In order to see the difference, 
we investigate two typical cases; Case I corresponding to $\alpha=-0.33$, 
$\delta=0.55$, $\xi=0$, which gives $\beta = 0.01$, $\gamma = 0.94$, $k = -0.10$, and 
it satisfies $\beta+ 3 \alpha k + \xi k > 0$. 
Case II corresponding to $\alpha=0.2$, 
$\delta=0.55$, $\xi=0.4$, which gives $\beta = 0.12$, $\gamma = 0.82$, $k = -0.33$, and 
it satisfes $\beta+ 3 \alpha k + \xi k < 0$. 

It is easily checked that for the case I, it  
satisfies eq.~(\ref{alphadeltaconditionone}), and therefore, (\ref{firstcondition}) - (\ref{thisisthesecondone}). 
So case I corresponds to the case where axion term is expected to be enhanced at low temperature $r_h \to 0$ due to $c_1 Q_e/r_h^{\beta+ 3 \alpha k + \xi k} \to \infty$, and we expect  that axion term induces instability even at very small values of $c_1$ at low temperature. 

On the other hand, 
for the case II, it  
violates 
eq.~(\ref{alphadeltaconditionone}) (more precisely it violates eq.~(\ref{thisisthesecondone}) since $\xi \neq 0$),  
but  does not violate the UV AdS$_4$ BF bound (\ref{deltasquare38forBF}).  Therefore case II corresponds to the case where axion term is expected to be rather suppressed at low temperature $r_h \to 0$ due to $c_1 Q_e/r_h^{\beta+ 3 \alpha k + \xi k} \to 0$, and we expect that we need large values of $c_1$ in order to induce instability at low temperature. 

For both cases, we first numerically construct the background solutions interpolating the 
IR analytical hyperscaling violating geometries and UV AdS$_4$ as we reviewed in \S \ref{section2B2}. 
Then we solve the coupled eq.~(\ref{coupledgaugeaxionone}) and eq.~(\ref{coupledgaugeaxiontwo})  and find numerically the solutions $\psi_\pm$ satisfying 
the boundary conditions~eq.~(\ref{bc_infinity}) for a suitable $c_1$, $q$ and $\psi_-$ 
for each temperature $T$ ($r_h$), where eq.~(\ref{Thinregardstorh}) gives the relationship between 
$r_h$ and $T$. 

For each temperature  
we find the zero mode solution by tuning $q$ 
for $c_1$, as far as $c_1 > c_{min}$. 
There is a minimum value $c_{min}$ for $c_1$, namely,  
we could not find any zero mode solution for any $q$ and $\psi_-$ 
when $c_1 < c_{min}$.  

We plot $c_{min}$ for each temperature $T$ 
in Fig.~4~(case I) and Fig.~5~(case II). 
\begin{figure}
 \begin{center}
  \includegraphics[width=8.2truecm,clip]{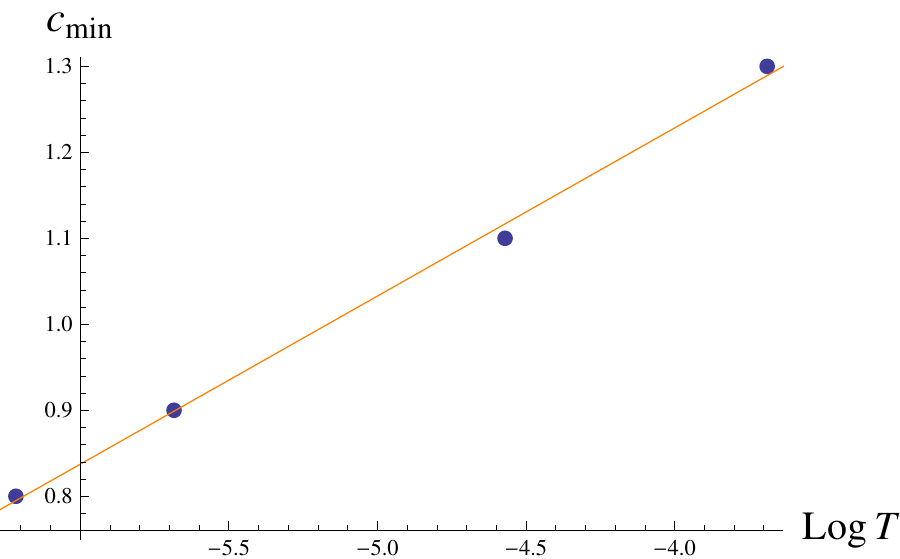}
  \caption{The plot of $c_{min}$ for various 
temperatures $T$ for 
case I~($\alpha=-0.33$, 
$\delta=0.55$, $\xi=0$). Each plot corresponds to ($c_{min}$, 
$T$, $q$) = ($1.3, \,2.05*10^{-4}, \,0.83$), \, 
($1.1, \,2.69*10^{-5}, \,0.85$), \, 
($0.9, \,2.07*10^{-6}, \,0.86$), \, 
($0.8, \,6.08*10^{-7}, \,1.0$). The parameters are approximately on the line $c_{min}(T) = 
2.0
+ 0.20 \log_{10} T$ in the temperature range we study.} 
 \end{center}
\end{figure}
\begin{figure}
 \begin{center}
  \includegraphics[width=8.2truecm,clip]{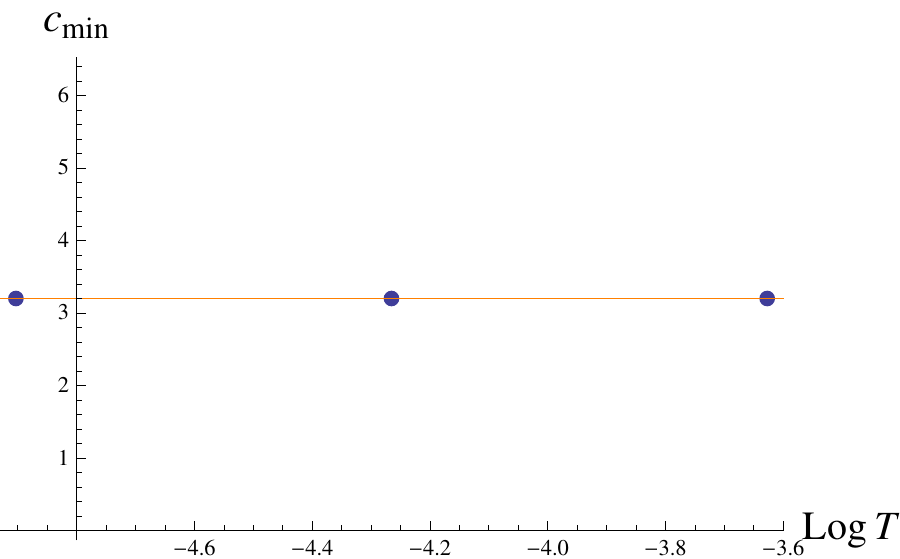}
  \caption{The plot of $c_{min}$ for various temperatures $T$ for 
case II~($\alpha=0.2$, $\delta=0.55$, $\xi=0.4$). Each plot corresponds 
to ($c_{min}$, 
$T$, $q$) = ($3.2, \,2.36*10^{-4}, \,0.45$), \, 
($3.2, \,5.43*10^{-5}, \,0.45$),\,
($3.2, \,1.25*10^{-5}, \,0.45$). $c_{min}(T)$ is almost constant,  $c_{min}(T) \approx 3.2$ in the 
temperature range we study.} 
 \end{center}
\end{figure}
%
As expected, for the case I, $c_{min}$ decreases significantly 
as $T$ decreases and seems to approach zero. 
On the other hand, for the case II, $c_{min}$ does not change drastically 
as $T$ decreases.  
This suggests that spontaneous translational symmetry breaking 
easily occurs for much lower temperature when eq.~(\ref{alphadeltaconditionone}) 
is satisfied, although we cannot further evaluate the minimum value 
$c_{min}$  
for much lower temperature. This is because highly numerical accuracy is required. 
Similarly we face another difficulties at $\log_{10} T > -3.6$. 
This is because in such temperature, $r_h$ is not so small and as a result, $\phi|_{r = r_h}$ is not large 
enough. Then, our perturbative method in \S \ref{section2B2} to construct full interpolating solutions breaks down. 
Therefore we conduct numerical analysis in rather restricted low temperature range, $-6.3 \lesssim \log_{10} T \lesssim -3.6$ 
for case I, and $-4.9 \lesssim \log_{10} T \lesssim -3.6$ 
for case II.

Note that the value range of $c_{min}$ for case I in Fig.~4 is  {\it lower}  than the critical value for the AdS$_2$ case, $c_{min \, AdS_2} = \sqrt{2}$, which we obtained in eq.~(\ref{cminads2case}). 
%
%
This result is consistent with the analysis in \S \ref{section3B} and \S \ref{section3C1} that 
these are instabilities triggered by the enhanced axion term effect due to the radius dependent factor, $c_1 Q_e/r_h^{\beta+ 3 \alpha k + \xi k} \to \infty$ at $r_h \to 0$.  
One might expect that in case I, at the zero temperature limit where $r_h \to 0$, 
the critical value $c_{min}$ approaches zero. Fig.~4 is consistent with this expectation. However due to the 
numerical difficulties, we could not confirm this at the very low temperature $T \lesssim 6 * 10^{-7}$. 
 

On the other hand, in case II, we find that the critical value $c_{min}$ is almost constant as 
we lower the temperature $r_h$. Note that the value range of $c_{min}$ for case II in Fig.~5 is 
 {\it higher}  than the critical value for the AdS$_2$ case, suggesting that 
these are instabilities triggered by the suppressed axion term effect due to the radius dependent factor 
in case II.   
However we do not have clear physical interpretation 
of the result in the case II. From the IR analysis, it might suggest 
that the critical value $c_{min}$ increases as we lower the temperature, but the result of Fig.~5 is not so. 
One possible reason for this behavior is that 
the negative momentum square is more dominating away from the horizon, $r \gg r_h$.  
As we have discussed in \S \ref{section3C1}, one can see that 
even in the case II with zero temperature limit, 
one of the eigenvalue of $M^2$ can become negative at some finite radius $r$. 
In other words, one of the eigenvalue of $M^2$ can become negative at some finite radius $r$, 
but as $r \to 0$, that value approaches zero. 
Because of this, even though one of the eigenvalue of $M^2$ becomes zero at the horizon in the case II, 
it can become some negative value at some finite radius $r$ and therefore, 
there could exist a zero mode in the whole bulk for the stripe instability.  
In such case, if the bulk region, where $M^2$ eigenvalue becomes negative, is away from the horizon $r_h$,   
then it is possible that lowering the temperature does not influence these bulk region 
much. As a result, in such a case, changing the $r_h$ does not change the $c_{min}$. 
However in order to obtain clear physical understanding of these results, we need more detail analysis.

It is also useful to draw the figure for the zero mode in $(q, T)$ plot with fixed $c_1$ value.  
\footnote{We thank Aristos Donos and Jerome Gauntlett for raising this question to us.}. 
We plot the critical temperatures $T$ versus $q$  
for the normalizable zero mode 
in Fig.~6~(case I parameter choice with $c_1 = 1.3$) and Fig.~7~(case II parameter choice with $c_1 = 3.3$). 
\begin{figure}
 \begin{center}
 \includegraphics[scale=.85]{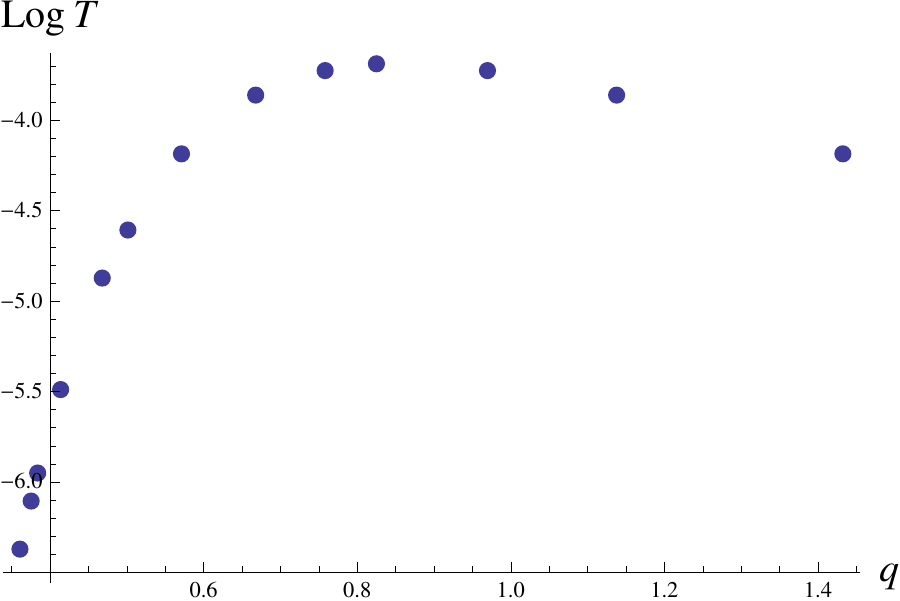}
  \caption{$T-q$ curve for the normalizable zero mode in full geometries for the case I parameter choice   
  ($\alpha = -0.33$, $\delta = 0.55$, $\xi = 0$), with $c_1 = 1.3$. 
  Due to the ``enhancement factor'' $1/r_h^{\beta + 3 \alpha k + \xi k} \approx 8$, even though $c_1$ is smaller  than the critical value for IR AdS$_2$ case, $c_{min \, AdS_2} = \sqrt{2}$, we have two $q$'s for the 
  zero model, and in between, there should be unstable modes. This figure is in good comparison with IR AdS$_2$ case analysis done in \cite{Donosstripes}, Figure 2, where, $c_1 > c_{min \, AdS_2}$.}
 \end{center}
\end{figure}

\begin{figure}
 \begin{center}
 \includegraphics[scale=.85]{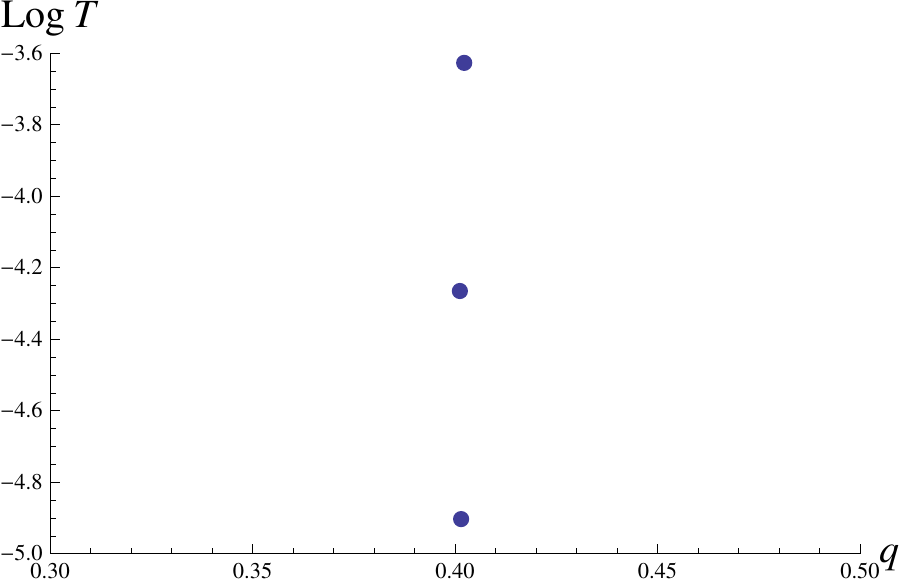}
  \caption{$T-q$ curve for the normalizable zero mode for the case II parameter choice ($\alpha = 0.2$, $\delta = 0.55$, $\xi = 4$), with $c_1 = 3.3$. }
 \end{center}
\end{figure}

In the case I, for given temperature, we generically have two $q$'s allowing the normalizable zero modes. 
Unstable modes should exist for momentum $q$ between $q_{\min}(T)$ and $q_{max}(T)$, and 
at $T \approx 2.05 * 10^{-4}$, $q_{min}$ and $q_{max}$ coincides at $q \approx 0.83$.   
As we lower the temperature, $q_{min}(T)$ decreases and $q_{max}(T)$ increases.   
In the case I, because $c_1 Q_e/r_h^{\beta + 3 \alpha k + \xi k}$ goes to infinity at $r_h \to 0$, 
we expect to have very large effect of the axion term, and therefore 
we expect that $q_{max}$ becomes very large, at zero temperature limit. 
It would be really nice to confirm this numerically, however we could not conduct numerics at this very low temperature, $T \lesssim 4.2 *10^{-7}$ due to the difficulties of numerical analysis.  

Again in the case II, we do not have clear physical interpretation of the results.

\section{Perturbation analysis beyond the probe limit}
\label{section4}

We have analyzed so far the system without the graviton fluctuation, namely in the 
probe limit. However, 
it is pretty straightforward to conduct the similar analysis with the graviton fluctuation, and  
we will see that the analysis with graviton fluctuation shows essentially the same results,   
compared with the analysis without graviton.   
We will see here that there is a negative momentum squared mode 
in the hyperscaling violating geometry even after we take into account the coupling to the graviton. 

We consider again the action (\ref{actionofourmodel}). 
The fluctuations we consider are the following components,  $ \delta g_{ty}$, 
$\delta A_y$, $\delta a$, $\delta \phi$, and 
we take the following mode dependence, 
\bea
\delta g_{ty} &=& \delta g_{ty}(r) \sin q x \,, \\\ 
\delta A_y &=& \delta A_y(r) \sin q x  \,, \quad \\ 
\delta a &=& \delta a(r) \cos q x \,, \quad \\
\delta \phi &=& 0 \,.
\eea
which has additional graviton mode. Here we have set $\omega = 0$, {\it i.e.}, 
no time-dependence from the beginning compared with eq.~(\ref{nogravitonflucAya}) and 
(\ref{nogravitonfluca}), in order 
to discuss the onset of the instability.  

Quite analogously to the previous analysis in \S \ref{section3A} in the probe limit, 
given the background geometry eq.~(\ref{metricansatz}), 
we have the equations of motion for the gauge field $\delta A_y$, 
\bea
\label{Ayequationofmotionwithgraviton}
&& - f(\phi)   b^{-2} q^2  \delta A_y(r) \nonumber \\
&& 
+ \partial_r \Bigl(  f(\phi) \tilde a^2   \partial_r  \delta A_y(r) 
+  f(\phi)   (\partial_r A_{t}) \delta g_{ty} (r)  \Bigr)  \nonumber \\
&& + 
(\partial_a \theta(a)) \frac{q Q_e}{b^2 f(\phi)}   \delta a(r) = 0 \,, \quad
\label{deltaAyequationwithgraviotnfluc}
\eea
and for the axion,  
\bea
\label{axionequationofmotionwithgraviton}
&&   \partial_r (e^{2 \xi \phi} \sqrt{-g} g^{rr} \partial_r \delta a(r) ) 
-  q^2 e^{2 \xi \phi} \sqrt{-g}    g^{xx} \delta a(r)  \nonumber \\
&&+ q 
 ( \partial_a \theta(a) )  F_{tr}  \delta A_{y}(r) = 0 \,. \quad  \, 
\eea
This equation is unmodified by $\delta g_{ty} \neq 0$. 

In addition, from the $(t,r)$ component of the trace reversed Einstein equations, 
we have fluctuation equation for the graviton, 
\bea
\label{deltagtyflucteq}
&& \tilde a^2 b^2
   \left( (\partial_r^2 \delta g_{ty}(r))- 4 f(\phi ) F_{tr}(\partial_r \delta A_y(r)) \right) \nonumber \\
&& + \Bigl(4 \tilde a  b  (\partial_r \tilde a)  (\partial_r b)  +b^2 \left(2 f(\phi ) (F_{tr})^2+V(\phi )\right) \nonumber \\
&& \quad  -q^2\Bigr)  \delta g_{ty}(r)  
 = 0  \,, 
\eea
and all the other components of  
Einstein equations and equations of motion are automatically satisfied. 

Let's investigate the near horizon limit in similar way to the analysis of \S \ref{section3B} 
and \S \ref{section3C1}. 
From graviton fluctuation eq.~(\ref{deltagtyflucteq}), by imposing the regularity condition of the solution at the horizon,  
we can see that we need the boundary condition $\delta g_{ty}(r = r_h) = 0$ at the horizon. 
Then, it is more convenient to set the new variable $\delta h_{ty} = \partial_r \delta g_{ty}$. 
And quite analogously to the case where we neglect graviton fluctuation in \S \ref{section3C1},    
the three equations eq.~(\ref{deltaAyequationwithgraviotnfluc}) - (\ref{deltagtyflucteq}) are approximated and written 
in the near horizon as   
\bea
\hat \nabla^2 \delta \tilde Y |_{r=r_h}  \approx \tilde M^2 |_{r=r_h}   \delta \tilde Y |_{r=r_h} 
\eea
where
\bea
\hat \nabla^2  &=& \partial_r   \tilde a^2 \partial_r   \,, \quad 
\delta \tilde Y = \left(   \begin{array}{cc}  { {\delta  A_y}}  \\  
{ {\delta a}} \\ {{\partial_r (\delta g_{ty}) }} \end{array} \right) \,, \,   
\eea
and 
\bea
%
%
%
%
%
\tilde M^2  \equiv 
\left(\begin{array}{ccc} 
{q^2}/{b^2} & 
(M_{12}(r))^2
& F_{tr}\\
(M_{21}(r))^2 
& {q^2}/{b^2} & 0 \\
 {4 f(\phi) F_{tr} q^2}/{b^2}   &
(M_{32}(r))^2
  & (M_{33}(r))^2 
\end{array} \right) ,  \nonumber \\ 
\eea
\bea
(M_{12}(r))^2 &=& 
-{( \partial_a \theta(a) )  q F_{tr}}/{f(\phi)} \,, \\
(M_{21}(r))^2 &=& 
{ -( \partial_a \theta(a) )  q F_{tr}}/{e^{2 \xi \phi} b^2} \,, \\
(M_{32}(r))^2 &=& 
  -4 (\partial_a \theta(a)) { q (F_{tr})^2} \,, \\
(M_{33}(r))^2 &=&  \frac{q^2}{b^2}  +  2 f(\phi ) (F_{tr})^2  
- \frac{4 a  b  (\partial_r a)  (\partial_r b)}{b^2} - V(\phi ) 
\,,\nonumber \\
\eea
and $\tilde M^2 |_{r=r_h}$ means $\tilde M^2$ 
evaluated at the horizon $r =r_h$ \footnote{ In order to derive this result, we have used two conditions; 
1)  terms proportional to $\delta g_{ty}$, are neglected 
since $\delta g_{ty}|_{r = r_h} \to 0$, and 
2)  using the regularity of the flux $F_{tr}$ and 
dilaton $f(\phi)$ at the finite temperature horizon $r = r_h$, 
we can neglect term proportional to $ \partial_r \delta A_y(r) $ since 
$| \bigl[ (\partial_r (f(\phi ) F_{tr}))  {\tilde a^2 }  \partial_r \delta A_y(r) \bigr] |_{r = r_h}| $ 
$\ll$ $| \bigl[ f(\phi ) F_{tr} ( \partial_r  {\tilde a^2 } ) \partial_r \delta A_y(r) \bigr] |_{r = r_h} |$ 
holds. }. 

The determinant of $\tilde M^2$ gives 
\bea
&& \mbox{Det} \tilde M^2 \nonumber \\
&& = 4  ( \partial_a \theta(a) )^2  q^2 (F_{tr})^4 {e^{-2 \xi \phi} b^{-2}} \nonumber \\
&& \quad -  ( \partial_a \theta(a) )^2  q^2 (F_{tr})^2 {f^{-1}(\phi)e^{-2 \xi \phi} b^{-2}} (M_{33})^2 \nonumber \\
&& \quad -  {4 f(\phi) (F_{tr})^2 q^4}{b^{-4}} + q^4 b^{-4} (M_{33})^2 \, \nonumber \\
&& = \Bigl(q^4  -  \frac{{ ( \partial_a \theta(a) )^2 q^2 b^2 (F_{tr}})^2 }{f(\phi)e^{2 \xi \phi} } 
 \Bigr) \nonumber \\
&& \quad \times \,  b^{-4}  \left((M_{33})^2 -4 f(\phi) (F_{tr})^2 \right) \,.
\eea
From eq.~(\ref{case1}) and (\ref{case2}), we can obtain the relation 
\bea
 - 2 \alpha k - 4 \beta = 2 \delta k = 2 \gamma -2 \,
\eea 
in the near horizon region of geometries $r \to r_h$ with hyperscaling violation, 
where $V(\phi) \to V_0 e^{2 \delta \phi}$, then using eq.~(\ref{explicitexamples}), (\ref{fluxansatz})  - (\ref{finiteTb}), 
we have   
\bea
&& \mbox{Det} \tilde M^2 |_{r = r_h}  \nonumber \\
&& =   r_h^{-6 \beta} \Bigl(q^4  -  {q^2 c_1^2    Q_e^2 r_h^{-2 \left( \beta + k (3 \alpha  + \xi )\right)}}  
 \Bigr) \nonumber \\
&& \quad \quad \quad \times \,   \left( q^2   - \zeta r_h^{2 \gamma + 2 \beta -2} \right)   \,,
\eea
where
\bea
&& \zeta \equiv 2 (Q_e)^2 + 2 (2 \beta + 2 \gamma - 1)\beta C_a^2  +  V_0 \,. 
\eea

Therefore, if 
\bea
\label{thisistheconditionwithgraviton}
\beta + k (3 \alpha + \xi) > 0  \,,
\eea
is satisfied, 
then the axion term proportional to $c_1 Q_e$ dominates. 
This is exactly the same condition, eq.~(\ref{finiteTcriteriaforeffectivemomentumenhancement}) and eq.~(\ref{thisisthesecondone}), which 
we have obtained in \S \ref{section3B} and \S \ref{section3C1} in the probe limit. 
In above $\mbox{Det}\, \tilde M^2$, gauge boson fluctuation $\delta A_y$ and axion fluctuation $\delta a$ gives the 
factor proportional to $(q^4  -  {q^2 c_1^2    Q_e^2 r_h^{-2 \left( \beta + k (3 \alpha  + \xi )\right)}})$. 
Note that this is $q^2_{eff +} \times q^2_{eff -}$ defined by eq.~(\ref{negativemasssquare}). And it is essentially 
$\mbox{Det}\, M^2$ given in eq.~(\ref{thedefinitionofmatrixm2forABCD}) - (\ref{thedefinitionofmatrixm2forD}), with eq.~(\ref{firstcondition}) and (\ref{secondcondition}) at $ r \to 0$ limit, up to overall factor 
$r^{2 (2 - 2 \beta - 2 \gamma)}$. Therefore even with the graviton fluctuation, $\delta A_y$ and $\delta a$ 
give the same mode for the instability. 

The addition of graviton fluctuation, 
simply adds one more eigenvalue to above $\mbox{Det} \, \tilde M^2$, and 
that eigenvalue is proportional to $( q^2   - \zeta r_h^{2 \gamma + 2 \beta -2})$. Therefore, 
even with the graviton fluctuation, the existence criteria of the negative eigenvalue mode of the matrix 
$\mbox{Det}\, \tilde M^2$ due to the axion term is precisely the same as the case without graviton fluctuation; 
We expect that if (\ref{thisistheconditionwithgraviton}) is satisfied, the effect of axion term 
are enhanced as we lower the temperature and intrigues more stripe instability, and 
the behavior of minimum $c_1$ for stripe instability is expected to show very similar behavior to the
 Fig.~4. On the other hand, if (\ref{thisistheconditionwithgraviton}) is not satisfied, we expect that 
 the effect of axion term are suppressed as we lower the temperature at the horizon 
 and this intrigues less instability, and 
the behavior of minimum $c_1$ for stripe instability is expected to show very similar behavior to the
 Fig.~5.
It would be best if we can confirm this by solving 
the eq.~(\ref{Ayequationofmotionwithgraviton}) - (\ref{deltagtyflucteq}) numerically with the normalizable 
boundary condition at the UV AdS$_4$ boundary as we have done in the probe limit in \S \ref{section3D}. 
However in this case, the parameter range we seek for the 
normalizable boundary conditions  becomes 3-dimensional, instead of 2-dimensional for the 
probe limit case, and this turns out quite hard task.  
Therefore we leave this as future work on this paper. 

Clearly at the large $c_1$ limit, we can 
have a mode which has large negative eigenvalue at some radius, and 
this indicates the striped phase instability can occur. 
One difference, compared to the probe limit, is that 
the analysis for the analytic expression for the onset of the instability 
in \S \ref{section3C2} does not work. This is because even if 
both eq.~(\ref{xiistuned}) and (\ref{betagammaequalone}) 
hold, the determinant of the matrix $\tilde M^2$ is proportional to $r^{-6 \beta}$. 
So we need $\beta=0$ and $\gamma = 1$ for the matrix $\tilde M^2$ to become constant matrix, which is AdS$_2$ case. This implies that we need to introduce one more parameter in the Lagrangian 
to be tuned, so that we can have constant matrix $\tilde M^2$ in the presence of graviton.


\section{
Summary and Discussion}
In this paper, we studied the stripe instabilities (spatially modulated instabilities) 
of the geometries with hyperscaling violation in the IR, which approach  AdS$_4$ metric in the UV 
asymptotically.  
The instabilities 
are induced by the axion term $\delta S =\int d^4x c_1 a F \wedge F$ in the bulk 4d action. 
We first study the perturbation equations in the probe limit, 
and saw that 
there is a strong correlation between the stripe instabilities caused by the axion term and parameters of the theories which determine the hyperscaling violation. 
Contrary to the IR AdS$_2$ case, 
we found that, due to the lack of scale invariance and the nontrivial radial dependence of the IR hyperscaling violating geometries, 
the effect of axion term can be either enhanced or suppressed depending on the 
parameters. 
In the parameter range where the effect of the axion term is expected to be enhanced, 
the stripe instability occurs and $c_{min}$ decreases as we lower the 
temperature, where $c_{min}$ is the critical value for the instability 
and instability occurs only at $c_1  \ge c_{min}$.   
On the other hand, in the parameter range where the effect of axion term is expected to be suppressed, we find that $c_{min}$ 
does not change much as we lower the temperature. 
We have explicitly obtained the zero mode solutions for the coupled fluctuations of gauge boson $\delta A_y$ and axion $\delta a$ numerically in the probe limit, with the boundary condition 
that there are no non-normalizable modes. 
This implies that in the dual field theories, 
the scalar and vector current $<j_y>$, which are dual to axion $a$ and gauge boson $A_y$ in the bulk, 
acquire the spatially modulated VEV 
spontaneously, 
and 
that 
dual theories at IR show the ``current density wave'' phase.  
We identify the instability onset on a certain one-parameter family of the hyperscaling violating geometries 
analytically, where the relation eq.~(\ref{betagammaequalone}) holds.  
We also argue that quite analogous results are expected to hold beyond the probe limit.

There are several open issues which should be understood in better way.  
We have done our search of the zero mode on rather limited temperature range in \S \ref{section3D}.  
This is due to the numerical difficulties, and  
it comes from the fact that our background solutions, which 
are hyperscaling violating geometries in IR and 
approach AdS$_4$ in UV, are constructed only numerically. If we could construct an analytical solution,  
we can search numerically for the zero mode more accurately.   
So it is interesting and important to look for analytical background 
solutions which interpolate between UV AdS$_4$ and IR hyperscaling violating geometries. 
By conducting the numerical analysis in better way, we can check if the 
$c_{min}$ goes to zero or not in the zero temperature limit in Fig.~4. 
Similarly, we can check how the $T-q$ curve behaves in the zero temperature limit in Fig.~6. 
It is interesting to check these.  

We argue in \S \ref{section4} that the results of instability analysis are essentially the same 
by taking into account the graviton effects.   
Of course, it is better if we could confirm this by searching for the zero mode explicitly on the full geometries, 
as we have done in \S \ref{section3D} in the probe limit.

There are other issues which we would like to understand in better way. 
In this paper, we did not argue the validity of the action eq.~(\ref{actionofourmodel}) 
for the background solutions. However, it can happens that 
the starting action is not valid for describing the solutions, depending on the 
behavior of the solutions. 
For example, our background dilaton has run-away behavior, and 
if the background dilaton runs to the strong coupling direction in IR, then we have to worry about 
the possibility that our starting action is highly corrected due to the strong coupling effects. 
Such a possibility exists if the starting action is derived under 
the weak coupling approximation. 
For example, from our action eq.~(\ref{actionofourmodel}), 
the effective coupling of the gauge field $g_{U(1)}$ is given by $g_{U(1)}^{-2} \equiv f(\phi) = e^{2 \alpha \phi}$. 
In our background solution 
this behaves as  $g_{U(1)} \to r^{-  \alpha k}$ in IR. If we assume that our theory eq.~(\ref{actionofourmodel}) 
is derived under the weak coupling condition, this with eq.~(\ref{case2}) forces us $- \alpha k \propto \alpha (\alpha + \delta) > 0$ for consistency. But this implies that our condition eq.~(\ref{5alphaminusdelta}) 
and (\ref{3alphaplusdelta}),  
for the case $\xi = 0$, cannot be satisfied. And we have only the parameter range 
where the axion term is expected to be suppressed, 
which corresponds to case II in the analysis of \S \ref{section3D}. 
These issues should be understood more from the string theory embedding view point. 
However rather in this paper, we study the stability analysis with the assumption that the action eq.~(\ref{actionofourmodel}) is valid for any solutions,  
we do not ``derive'' our action eq.~(\ref{actionofourmodel}) from string theory. 
It would be nice to study these consistency points in more detail. 

It could be that if the background dilaton blows up in the IR hyperscaling violating geometries, then 
we need to take into account the higher loop corrections and 
this might make the geometries into 
the AdS$_2$ metric in further IR, as studied, for examples, in \cite{Harrison:2012vy, Bhattacharya:2012zu, Kundu:2012jn}. This viewpoint resolves the problematic singularities of the 
zero temperature limit of the hyperscaling violating geometries at $r \to 0$ \footnote{This 
singularities can be avoided by introducing small but non-zero temperature. 
And our studies of the stripe instability 
in small but non-zero temperature case are not affected by this singularities.}.   
However, it is not clear if this is always the case. For examples,  once higher loop corrections 
(corresponds to higher string coupling $g_s$ corrections) enter the game, 
we always need to worry about full loop correction effects. 
Namely, once we face the situations where higher order $g_s$ effects are as important as leading 
order in $g_s$ expansion, 
this implies that $g_s$ expansion is no more valid. 
But in general, we do not have such a fully 
non-perturbative effective action, and the validity of the loop corrected effective action in order to derive the 
deep IR AdS$_2$ metric is unclear. 

Another interesting question is the end point of the stripe instability. 
In this paper, we studied the onset of the stripe instabilities. However 
to see what is the end point of these instabilities, perturbation analysis is not enough.
We need to study the equations of motion where $A_y$ and $a$ are coupled in the probe limit, 
and the full back reacted Einstein equations to go beyond the probe limit.  
For successful examples of the end point of stripe instabilities, see \cite{ooguri, gauntlett, Rozali}.  

In the probe limit, we identify the onset of the stripe instability when   
the relation eq.~(\ref{betagammaequalone}) holds. 
It is interesting to note that this relation also holds at the transition point 
between quasi-particle picture holds/breaks down from the study of the 
fermion Green's function on these background \cite{Iizuka:2011hg}. It is interesting to investigate to see 
if there are any deep reason for this coincident. 

There are many open questions to be understood better. However one thing which is very clear is that
these geometries with hyperscaling violation and stripe instabilities are rich subject and it is  
worth understanding more in great detail. 
We hope to return these questions in future. 
$\\$

{\bf Note added:} 
When we have almost finished preparing for the draft, 
a paper appeared \cite{Cremonini:2012ir}, where they also studied the 
stripe instability on the hyperscaling violating geometries. 
However our set-up and analysis is 
different from the one in \cite{Cremonini:2012ir}. 
The authors of \cite{Cremonini:2012ir} studied the geometries whose IR ($r \to 0$) is   
AdS$_2$, and in large $r$, they approach the hyperscaling violating geometries.  
They identified the onset of stripe instability in this IR AdS$_2$ region by using the AdS$_2$ BF bound, 
{\it i.e.,} local AdS$_2$ argument. 
Then they interpret that onset in terms of the large $r$ hyperscaling violating parameters. 
On the other hand, in this paper we studied 
 the geometries whose IR are hyperscaling violating geometries, which interpolate to the AdS$_4$ in the UV.  

\acknowledgments
We would like to thank Sera Cremonini, Aristomenis Donos, Jerome Gauntlett, Koji Hashimoto, and 
Annamaria Sinkovics for helpful discussion and  comments/questions on the draft.  
N.I. would like to thank ICTP South American Institute for Fundamental
Research for kind hospitality where part of this work was done. 
K.M. is supported in part by 
MEXT/JSPS KAKENHI Grant Number 23740200.


\appendix


\end{document}